\documentclass[twocolumn,showpacs,preprintnumbers,amsmath,amssymb]{revtex4}

\usepackage{graphicx}% Include figure files
\usepackage{bm}% bold math
\usepackage[dvips]{color}

\newcommand{\greysquare}{\color[rgb]{0.4,0.4,0.4}\blacksquare\color{black}}
\input epsf
\begin{document}

\title{Coarse-graining of cellular automata, emergence,  and the
predictability of complex systems}
\author{Navot Israeli}
\email{navot@walesltd.co.il}
\affiliation{Department of Physics of Complex Systems, Weizmann
Institute of Science, Rehovot, 76100, Israel.}
\author{Nigel Goldenfeld}
\email{nigel@uiuc.edu}
\affiliation{Department of Physics, University of Illinois at
Urbana-Champaign, 1110 West Green Street, Urbana, Illinois, 61801-3080.}
%\date{\today}

\begin{abstract}
We study the predictability of emergent phenomena in complex systems.
Using nearest neighbor, one-dimensional Cellular Automata (CA) as an
example, we show how to construct local coarse-grained descriptions of
CA in all classes of Wolfram's classification. The resulting
coarse-grained CA that we construct are capable of emulating the
large-scale behavior of the original systems without accounting for
small-scale details. Several CA that can be coarse-grained by this
construction are known to be universal Turing machines; they can
emulate any CA or other computing devices and are therefore
undecidable. We thus show that because in practice one only seeks
coarse-grained information, complex physical systems can be predictable
and even decidable at some level of description. The renormalization
group flows that we construct induce a hierarchy of CA rules. This
hierarchy agrees well with apparent rule complexity and is therefore a
good candidate for a complexity measure and a classification method.
Finally we argue that the large scale dynamics of CA can be very
simple, at least when measured by the Kolmogorov complexity of the
large scale update rule, and moreover exhibits a novel scaling law. We
show that because of this large-scale simplicity, the probability of
finding a coarse-grained description of CA approaches unity as one goes
to increasingly coarser scales. We interpret this large scale
simplicity as a pattern formation mechanism in which large scale
patterns are forced upon the system by the simplicity of the rules that
govern the large scale dynamics.
\end{abstract}

%47.54.+r Pattern selection; pattern formation
%05.45.Ra Coupled map lattices
%05.10.Cc Renormalization group methods
%05.20.-y Classical statistical mechanics
%05.45.-a Nonlinear dynamics and nonlinear dynamical systems (see also section 45 Classical mechanics of discrete systems)
\pacs{05.45.-a, 05.10.Cc, 47.54.+r}
\maketitle

\section{Introduction}
The scope of the growing field of \lq\lq complexity science" (or \lq\lq
complex systems") includes a broad variety of problems belonging to
different scientific areas. Examples for \lq\lq complex systems" can be
found in physics, biology, computer science, ecology, economy,
sociology and other fields. A recurring theme in most of what is
classified as \lq\lq complex systems" is that of {\it emergence}.
Emergent properties are those which arise spontaneously from the
collective dynamics of a large assemblage of interacting parts. A basic
question one asks in this context is how to derive and predict the
emergent properties from the behavior of the individual parts.  In
other words, the central issue is how to extract large-scale, global
properties from the underlying or microscopic degrees of freedom.

In the physical sciences, there are many examples of emergent phenomena
where it is indeed possible to relate the microscopic and macroscopic
worlds.  Physical systems are typically described in terms of
equations of motion of a huge number of microscopic degrees of freedom
(e.g. atoms). The microscopic dynamics is often erratic and complex,
yet in many cases it gives rise to patterns with characteristic length
and time scales much larger than the microscopic ones (e.g. the
pressure and temperature of a gas). These large scale patterns often
posses the interesting, physically relevant properties of the system
and one would like to model them or simulate their behavior. An
important problem in physics is therefore to understand and predict the
emergence of large scale behavior in a system, starting from its
microscopic description. This problem is a fundamental one because most
physical systems contain too many parts to be simulated directly and
would become intractable without a large reduction in the number of
degrees of freedom. A useful way to address this issue is to construct
coarse-grained models, which treat the dynamics of the large scale
patterns. The derivation of coarse-grained models from the microscopic
dynamics is far from trivial. In most cases it is done in a
phenomenological manner by introducing various (often uncontrolled)
approximations.

The problem of predicting emergent properties is most severe in systems
which are modelled or described by {\it undecidable} mathematical
algorithms\cite{wolf3,moore90}. For such systems there exists no
computationally efficient way of predicting their long time evolution.
In order to know the system's state after (e.g.) one million time steps
one must evolve the system a million time steps or perform a
computation of equivalent complexity. Wolfram has termed such
systems {\it computationally irreducible} and suggested that their
existence in nature is at the root of our apparent inability to model
and understand complex systems \cite{wolf3,wolf5,nks,ilachinski}. It is
tempting to conclude from this that the enterprise of physics itself is
doomed from the outset; rather than attempting to construct solvable
mathematical models of physical processes, computational models should
be built, explored and empirically analyzed.  This argument, however,
assumes that infinite precision is required for the prediction of
future evolution. As we mentioned above, usually coarse-grained or even
statistical information is sufficient. An interesting question that
arises is therefore: is it possible to derive coarse-grained models of
undecidable systems and can these coarse-grained models be decidable
and predictable?

In this work we address the emergence of large scale patterns in
complex systems and the associated predictability problems by studying
Cellular-Automata (CA). CA are spatially and temporally discrete
dynamical systems composed of a lattice of cells. They were originally
introduced by von Neumann and Ulam \cite{vonneumann} in the 1940's as a
possible way of simulating self-reproduction in biological systems.
Since then, CA have attracted a great deal of interest in physics
\cite{wolf1,physicaD10and45,ilachinski,wolf_collected_papers} because
they capture two basic ingredients of many physical systems: 1) they
evolve according to a local uniform rule. 2) CA can exhibit rich
behavior even with very simple update rules. For similar and other
reasons, CA have also attracted attention in computer science
\cite{mitchell98,sarkar00}, biology \cite{eek93}, material science
\cite{raabe02} and many other fields.  For a review on the literature
on CA see Refs.\ \onlinecite{ilachinski,wolf_collected_papers,nks}.

The simple construction of CA makes them accessible to computational
theoretic research methods. Using these methods it is sometimes
possible to quantify the complexity of CA rules according to the types
of computations they are capable of performing. This together with the
fact that CA are caricatures of physical systems has led many authors
to use them as a conceptual vehicle for studying complexity and pattern
formation. In this work we adopt this approach and study the
predictability of emergent patterns in complex systems by attempting to
systematically coarse-grain CA. A brief preliminary report of our
project can be found in Ref.\ \onlinecite{navot04}.

There is no unique way to define coarse-graining, but here we will
mean that our information about the CA is locally coarse-grained
in the sense of being stroboscopic in time, but that nearby cells
are grouped into a supercell according to some specified rule (as
is frequently done in statistical physics). Below we shall
frequently drop the qualifier "local" whenever there is no cause
for confusion. A system which can be coarse-grained is {\it
compact-able} since it is possible to calculate its future time
evolution (or some coarse aspects of it) using a more compact
algorithm than its native description. Note that our use of the
term compact-able refers to the phase space reduction associated
with coarse-graining, and is agnostic as to whether or not the
coarse-grained system is decidable or undecidable. Accordingly, we
define {\it predictable} to mean that a system is decidable or has
a decidable coarse-graining. Thus, it is possible to calculate the
future time evolution of a predictable system (or some coarse
aspects of it) using an algorithm which is more compact than both
the native and coarse-grained descriptions.

Our work is organized as follows. In section \ref{ca_intro} we give an
introduction to CA and their use in the study of complexity. In section
\ref{cg_procedure} we present a procedure for coarse-graining CA.
Section \ref{cg_results} shows and discusses the results of applying
our procedure to one dimensional CA. Most of the CA that we attempt to
coarse-grain are Wolfram's 256 elementary rules for nearest-neighbor
CA. We will also consider a few other rules of special interest. In
section \ref{Kolmogorov_complexity} we consider whether the
coarse-grain-ability of many CA that we found in the elementary rule
family is a common property of CA. Using computational theoretic
arguments we argue that the large scale behavior of local processes
must be very simple. Almost all CA can therefore be coarse-grained if
we go to a large enough scale. Our results are summarized and discussed
in \ref{conclusions}.

\section{Cellular Automata}
\label{ca_intro} Cellular automata are a class of homogeneous,
local and fully discrete dynamical systems. A cellular automaton
$A=\left(a\left(t\right),\{S_A\},f_A\right)$ is composed of a
lattice $a(t)$ of cells that can each assume a value from a finite
alphabet $\{S_A\}$. We denote individual lattice cells by
$a_{\alpha}\left( t \right)$ where the indexing reflects the
dimensionality and geometry of the lattice. Cell values evolve in
discrete time steps according to the pre-prescribed update rule
$f_A$. The update rule determines a cell's new state as a function
of cell values in a finite neighborhood. For example, in the case
of a one dimensional, nearest-neighbor CA the update rule is a
function $f_A:\{S_A\}^3\rightarrow \{S_A\}$ and
$a_n\left(t+1\right)=f_A\left[a_{n-1}\left(t \right),a_{n}\left(t
\right),a_{n+1}\left(t \right)\right]$. At each time step, each
cell in the lattice applies the update rule and updates its state
accordingly. The application of the update rule is done in
parallel for all the cells and all the cells apply the same rule.
We denote the application of the update rule on the entire lattice
by $a\left(t+1\right)=f_A\cdot a\left( t\right)$.

In early work \cite{wolf2,wolf3,wolf5,nks}, Wolfram proposed that CA
can be grouped into four classes of complexity. Class 1 consists of CA
whose dynamics reaches a steady state regardless of the initial
conditions. Class 2 consists of CA whose long time evolution produces
periodic or nested structures. CA from both of these classes are simple
in the sense that their long time evolution can be deduced from running
the system a small number of time steps. On the other hand, class 3 and
class 4 consist of \lq\lq complex" CA. Class 3 CA produce structures
that seem random.  Class 4 CA produce localized structures that
propagate and interact in a complex way above a regular background.
This classification is heuristic and the assignment of CA to the four
classes is somewhat subjective. Successive works on CA attempted to
improve it or to find better
alternatives\cite{culik88,gutowitz88,gutowitz90,stuner90,binder91,braga95,jin03}.
To the best of our knowledge there is, to date, no universally agreed
upon classification scheme of CA.

Based on numerical experiments, Wolfram hypothesized that most of class
3 and 4 CA are {\it Computationally Irreducible}\cite{wolf2,wolf3,nks}.
Namely, the evolution of these CA cannot be predicted by a process
which is drastically more efficient than themselves. In order to
calculate the state of a Computationally Irreducible CA after $t$ time
steps, one must run the CA for $t$ time steps or perform a computation
of equivalent complexity. This definition is somewhat loose because it
is not always clear how to compare computation running times and
efficiency on different architectures. In addition, Wolfram recognized
that even computationally irreducible systems may have some \lq\lq
superficial reducibility" (see page 746 in Ref.\ \onlinecite{nks}) and
can be reduced to a limited extent. The difference between \lq\lq
superficial" and true reducibility however is not well defined. It is
nevertheless clear that the asymptotic $t\rightarrow \infty$ behavior
of a Computationally Irreducible system cannot be predicted by any
computation of finite size. Wolfram further argued that Computationally
Irreducible systems are abundant in nature and that this fact explains
our inability as physicists to deal with complex systems
\cite{wolf3,wolf5,nks,ilachinski}.

It is difficult in general to tell whether a CA, behaving in an
apparently complex way, is Computationally Irreducible. More concrete
properties of CA which are related to Computational Irreducibility are
{\it Undecidability} and {\it Universality}. Mathematical processes are
said to be undecidable when there can be no algorithm that is
guaranteed to predict their outcome in a finite time. Equivalently, CA
are said to be undecidable when aspects of their dynamics are
undecidable. Computationally Irreducible CA are therefore Undecidable
and in the weak asymptotic definition that we gave above, Computational
Irreducibility is equivalent to Undecidability. For lack of a better choice we adopt this asymptotic definition
and in the reminder of this work we will use the two terms
interchangeably.

Some CA are known to be universal Turing machines\cite{herken} and
are capable of performing all computations done by other
processes. A famous two dimensional example is Conway's game of
life\cite{gardner70}; several examples in one dimension are
Lindgren and Nordahl \cite{lindgren90}, Albert and Culik
\cite{albert87} and Wolfram's rule 110 \cite{nks}. Universal CA
are, in a sense, maximally complex because they can emulate the
dynamics of all other CA. Being universal Turing machines, these
CA are subject to undecidable questions regarding their
dynamics\cite{wolf2}. For example whether an initial state will
ever decay into a quiescent state is the CA equivalence of the
undecidable halting problem\cite{herken}. Universal CA are
therefore Undecidable.

Wolfram's classification of CA is topological in the sense that CA are
classified according to the properties of their trajectories. A
different, more ambitious, approach is to classify CA according to a
parameter derived directly from their rule tables. Langton
\cite{langton90} suggested that CA rules can be parameterized by his
$\lambda$ parameter which measures the fraction of non-quiescence rule
table entries. He showed a strong correlation between the value of
$\lambda$ and the complexity found in the CA trajectories. For small
values of $\lambda$ one characteristically finds class 1 and 2 behavior
while for $\lambda \sim 1$ a class 3 behavior is usually observed.
Langton identified a narrow region of intermediate values of $\lambda$
where he found class 4 characteristic behavior. Based on these
observations Langton proposed the {\it edge of chaos}
hypothesis\cite{langton90}. This hypothesis claims that in the space of
dynamical systems, interesting systems which are capable of computation
are located at the boundary between simple and chaotic systems. This
appealing hypothesis however was criticized in later works
\cite{mitchell93}. Recently, a different parametrization of CA rule
tables was proposed by Dubacq et al. \cite{dubacq01}. This new approach
is based on the information content of the rule table as measured by
its Kolmogorov Complexity. As we will show below, our results lend
support to this notion and indicate that rule tables with low
Kolmogorov complexity lead to simple behavior and vice versa.

In addition to attempts to find order and hierarchy in the space of CA
rules, much research has been devoted to the study of CA classes with
special properties. Additive CA (or linear)
\cite{robinson87,barbe95}, commuting CA \cite{voorhees93} and CA
with certain algebraic properties \cite{moore97,moore98} are a few
examples. Unsurprisingly, the dynamics of CA which enjoy such
special properties can in most cases be understood and predictable
at some level.

In this work we will mostly be concerned with the family of one
dimensional, nearest neighbor binary CA that were the subject of
Wolfram's investigations. These 256 elementary rules are among the
simplest imaginable CA and thus present us with the least
computational challenges when attempting to coarse-grain them. We
will use Wolfram's notation\cite{wolf1} for identifying individual
rules. The update function of an elementary rule is described by a
rule number between 0 and 255. The eight bit binary representation
of the rule number specifies the update function outcome for the
eight possible three cell configurations (where \lq\lq 000" is the least significant and \lq\lq 111" is the most significant bit). CA are often conveniently visualized with different colors denoting different cell values. When dealing with binary CA we will use the convention $\square=0$, $\blacksquare=1$ and use the two notations interchangeably.

\section{Local coarse-graining of Cellular Automata}
\label{cg_procedure} We now turn to study the emergence of large scale
patterns in CA and the associated predictability problems by attempting
to coarse-grain CA. There are many ways to define a coarse-graining of
a dynamical system. In this work we define it as a (real-space)
renormalization scheme where the original CA
$A=\left(a\left(t\right),\{S_A\},f_A\right)$ is coarse-grained to a
renormlized CA $B=\left(b\left(t\right),\{S_B\},f_B\right)$ through the
lattice transformation $b_k=P\left(a_{N\cdot k},a_{N\cdot
k+1},\dots,a_{N\cdot k+N-1}\right)$. The projection function
$P:\{S_A\}^N\rightarrow\{S_B\}$ projects the value of a block of $N$
cells in $A$, which we term a {\it supercell}, to a single cell in $B$.
By writing $P\cdot a$ we denote the block-wise application of $P$ on
the entire lattice $a$. Only non-trivial cases where $P$ is
irreversible are considered because we want $B$ to provide a partial
account of the full dynamics of $A$.

In order for $B$ and $P$ to provide a coarse-grained emulation of
$A$ they must satisfy the commutativity condition
\begin{equation}
P\cdot f_A^{T} \cdot a(0)=f_B\cdot P\cdot a(0)\;,
\label{coarse-graining_def}
\end{equation}
for every initial condition $a(0)$ of $A$. The constant $T$ in the
above equation is a time scale associated with the
coarse-graining. A repeated application of Eq.\
(\ref{coarse-graining_def}) shows that
\begin{equation}
P\cdot f_A^{T\cdot t} \cdot a(0)=f_B^{t}\cdot P\cdot a(0)\;,
\label{coarse-graining_def2}
\end{equation}
for all $t$. Namely, running the original CA for $T\cdot t$ time
steps and then projecting is equivalent to projecting the initial
condition and then running the renormalized CA for $t$ time steps.
Thus, if we are only interested in the projected information we
can run the more efficient CA $B$.

Renormalization group transformations in statistical physics are
usually performed with projection operators that arise from a physical
intuition and understanding of the system in question. Majority rules
and different types of averages are often the projection operators of
choice. In this work we have the advantage that the CA we wish to
coarse-grain are fully discrete systems and the number of possible
projections of a supercell of size $N$ is finite. We will therefore
consider all possible (at least with small supercells) projection
operators and will not restrict ourselves to coarse-graining by
averaging. In addition, the discrete nature of CA makes it very
difficult to find useful approximate solutions of Eq.\
(\ref{coarse-graining_def}) because there is no natural small parameter
that can be used to construct perturbative coarse-graining schemes. We
therefore require that Eq.\ (\ref{coarse-graining_def}) is satisfied
exactly.

\subsection{Coarse-graining procedure}
We now define a simple procedure for
coarse-graining CA. Other constructions are undoubtedly
possible.  For simplicity we limit our treatment to
one-dimensional systems with nearest neighbor interactions.
Generalizations to higher dimensions and different interaction
radii are straightforward.

The commutativity condition Eq.\ (\ref{coarse-graining_def})
implies that the renormlized CA $B$ is homomorphic to the dynamics of $A$ on the scale defined by the supercell size
$N$. To search for explicit coarse-graining rules, we define the
$N$'th supercell version
$A^N=\left(a^N,\{S_{A^N}\},f_{A^N}\right)$ of $A$. Each cell of
$A^N$ represents $N$ cells of $A$ and accepts values from the
alphabet $\{S_{A^N}\}=\{S_A\}^N$ which includes all possible
configurations of $N$ cells in $A$. The transition function
$f_{A^N}$ of the supercell CA can be defined in many ways
depending on our choice of the supercells interaction radius. Here
we choose $A^N$ to be a nearest neighbor CA and compute
$f_{A^N}:\left\{S_{A^N}\right\}^3\rightarrow
\left\{S_{A^N}\right\}$ by running $A$ for $N$ time steps on all
possible initial conditions of length $3N$. In this way $A^N$
follows the dynamics of $A$ and each application of $A^N$ computes
the evolution of a block of $N$ cells of $A$, for $N$ time steps.
This choice will later result in a coarse-grained CA $B$ which is
itself nearest-neighbor. This is convenient because it enables us
to compare the original and coarse-grained systems.
Another convenient feature of this construction is that it renders
the coarse-graining time scale $T$ equal to the supercell size
$N$. Other constructions however are undoubtedly possible. Note that
$A^N$ is not a coarse-graining of $A$ because no information was
lost in the cell translation.

Next we attempt to generate the coarse CA $B$ by projecting the
alphabet of $A^N$ on a subset $\{S_B\} \subset \{S_{A^N}\}$ which
will serve as the alphabet of $B$. This is the key step where
information is being lost. The transition function
$f_B$ is constructed from $f_{A^N}$ by projecting its arguments
and outcome:
\begin{equation}
f_B\left[P(x_1),P(x_2),P(x_3)\right]=
P\left(f_{A^N}\left[x_1,x_2,x_3\right]\right).
\label{fb_mv_construction}
\end{equation}
Here $P\left(x\right)$ denotes the projection of the supercell
value $x$. This construction is possible only if
\begin{eqnarray}
P\left(f_{A^N}\left[x_1,x_2,x_3\right]\right)&=&
P\left(f_{A^N}\left[y_1,y_2,y_3\right]\right), \nonumber \\
&&\forall \left(x,y|P(x_i)=P(y_i)\right).
\label{trans_func_projection}
\end{eqnarray}
Otherwise, $f_B$ is multi-valued and our
coarse-graining attempt fails for the specific choice of $N$ and $P$.

Equations (\ref{fb_mv_construction}) and (\ref{trans_func_projection}) can
also be cast in the matrix form
\begin{equation}
P\cdot A^N=B \cdot P_3\;,
\label{cg_matrix_form}
\end{equation}
which may be useful. Here $A^N$ is an $S_{A^N}\times (S_{A^N})^3$
matrix which specify the $N$ cell block output for every possible
combination of $3N$ cells. $P$ is an $S_B\times S_{A^N}$ matrix
that project from $S_{A^N}$ to $S_B$. $P_3$ is a $(S_B)^3\times
(S_{A^N})^3$ matrix which projects 3 consecutive super cells and
is a (simple) function of $P$. The coarse-grained CA $B$ is an
$S_B\times (S_B)^3$ matrix and is also a function of $P$. This is
a greatly over determined equation for the projection operator
$P$. For a given value of $N$ and $S_B$ the equation contains
$S_B\times (S_{A^N})^3$ constraints while $P$ is defined by
$S_{A^N}$ free parameters.

In cases where Eq.\ (\ref{trans_func_projection}) is satisfied, the resulting CA $B$ is a coarse-graining of $A^N$ with a time scale $T=1$.
For every step
$a^N_n(t+1)=f_{A^N}\left[a^N_{n-1}(t),a^N_n(t),a^N_{n+1}(t)\right]$ of
$A^N$, $B$ makes the move
\begin{eqnarray}
b_n(t&+&1)=f_B\left[b_{n-1}(t),b_n(t),b_{n+1}(t)\right] \\
&=&P\left(f_{A^N}\left[a^N_{n-1}(t),a^N_n(t),a^N_{n+1}(t)\right]\right) \nonumber \\
&=&P\left(a^N_n(t+1)\right)\;,
\nonumber
\end{eqnarray}
and therefore satisfies Eq.\ (\ref{coarse-graining_def}).
Since a single time step of $A^N$ computes $N$ time steps of $A$,
$B$ is also a coarse-graining of $A$ with a coarse-grained time
scale $T=N$. Analogies of these operators have been used in
attempts to reduce the computational complexity of certain
stochastic partial differential equations
\cite{hou01,degenhard02}. Similar ideas have been used to
calculate critical exponents in probabilistic CA
\cite{oliveira97,monetti98}.

To illustrate our method let us give a simple example. Rule 128 is
a class 1 elementary CA defined on the $\{\square,\blacksquare\}$
alphabet with the update function
\begin{eqnarray}
\lefteqn{f_{128}\left[x_{n-1},x_n,x_{n+1}\right]=} \nonumber \\
&&\left\{
\begin{array}{l}
\square\;,\;x_{n-1},x_n,x_{n+1}\neq
\blacksquare,\blacksquare,\blacksquare \\
\blacksquare\;,\;x_{n-1},x_n,x_{n+1}=\blacksquare,\blacksquare,\blacksquare\\
\end{array} 
\right. \;. \label{f128}
\end{eqnarray}
Figure \ref{cgof146figure} b) shows a typical evolution of this
simple rule where all black regions which are in contact with
white cells decay at a constant rate. To coarse-grain rule 128 we
choose a supercell size $N=2$ and calculate the supercell update
function
\begin{eqnarray}
\lefteqn{f_{128}^2\left[y_{n-1},y_n,y_{n+1}\right]=} \nonumber \\
&&\left\{
\begin{array}{l}
\blacksquare\blacksquare\;,\;y_{n-1},y_n,y_{n+1}=\blacksquare\blacksquare,\blacksquare\blacksquare,\blacksquare\blacksquare
\\
\square\blacksquare\;,\;y_{n-1},y_n,y_{n+1}=\square\blacksquare,\blacksquare\blacksquare,\blacksquare\blacksquare
\\
\blacksquare\square\;,\;y_{n-1},y_n,y_{n+1}=\blacksquare\blacksquare,\blacksquare\blacksquare,\blacksquare\square
\\
\square\square\;,\;\mbox{all other combinations}
\end{array}
\right.\;. \label{rule128supercellf}
\end{eqnarray}
Next we project the supercell alphabet using
\begin{equation}
P\left(y\right)=\left\{
\begin{array}{l}
\square\;,\;y\neq \blacksquare\blacksquare \\
\blacksquare\;,\;y=\blacksquare\blacksquare \\
\end{array}
\right.\;.
\end{equation}
Namely, the value of the coarse-grained cell is black only when
the supercell value corresponds to two black cells. Applying this
projection to the supercell update function Eq.\
(\ref{rule128supercellf}) we find that
\begin{eqnarray}
\lefteqn{P\left(f_{128}^2\left[P(y_{n-1}),P(y_n),P(y_{n+1})\right]\right)=} \nonumber \\
&&\left\{
\begin{array}{l}
\blacksquare\;,\;P(y_{n-1}),P(y_n),P(y_{n+1})=\blacksquare,\blacksquare,\blacksquare
\\
\square\;,\;P(y_{n-1}),P(y_n),P(y_{n+1})\neq
\blacksquare,\blacksquare,\blacksquare \\
\end{array}
\right. \;,
\end{eqnarray}
which is identical to the original update function $f_{128}$. Rule
128 can therefore be coarse-grained to itself, an expected result
due to the scale invariant behavior of this simple rule.

\subsection{Relevant and irrelevant degrees of freedom}
\label{r_and_ir_dof} It is interesting to notice that the above
coarse-graining procedure can lose two very different types of
dynamic information. To see this, consider Eq.\
(\ref{trans_func_projection}). This equation can be satisfied in
two ways. In the first case
\begin{eqnarray}
f_{A^N}\left[x_1,x_2,x_3\right]&=&f_{A^N}\left[y_1,
y_2,y_3\right],\nonumber \\
&&\forall \left(x,y|P(x_i)=P(y_i)\right)\;,
\label{irrelevant_cond}
\end{eqnarray}
which necessarily leads to Eq.\  (\ref{trans_func_projection}).
$f_{A^N}$ in this case is insensitive to the projection of its
arguments. The distinction between two variables which are identical
under projection is therefore {\it irrelevant} to the dynamics of
$A^N$, and by construction to the long time dynamics of $A$. By
eliminating irrelevant degrees of freedom (DOF), coarse-graining of
this type removes information which is redundant on the microscopic
scale. The coarse CA in this case accounts for all possible long time
trajectories of the original CA and the complexity classification of
the two CA is therefore the same.

In the second case Eq.\ (\ref{trans_func_projection}) is satisfied even
though Eq.\ (\ref{irrelevant_cond}) is violated.
Here the distinction between two variables which are identical under
projection is {\it relevant} to the dynamics of $A$. Replacing $x$ by
$y$ in the initial condition may give rise to a difference in the
dynamics of $A$. Moreover, the difference can be (and in many occasions
is) unbounded in space and time. Coarse-graining in this case is
possible because the difference is constrained in the cell state space by
the projection operator. Namely, projection of all such different
dynamics results in the same coarse-grained behavior. Note that the
coarse CA in this case cannot account for all possible long time
trajectories of the original one. It is therefore possible for the
original and coarse CA to fall into different complexity
classifications.

Coarse-graining by elimination of relevant DOF removes information
which is not redundant with respect to the original system. The
information becomes redundant only when moving to the coarse scale. In
fact, \lq\lq redundant" becomes a subjective qualifier here since it
depends on our choice of coarse description. In other words, it depends
on what aspects of the microscopic dynamics we want the coarse CA to
capture.

Let us illustrate the difference between coarse-graining of
relevant and irrelevant DOF. Consider a dynamical system whose
initial condition is in the vicinity of two limit cycles.
Depending on the initial condition, the system will flow to one of
the two cycles. Coarse-graining of irrelevant DOF can project all
the initial conditions on to two possible long time behaviors. Now
consider a system which is chaotic with two strange attractors.
Coarse-graining irrelevant DOF is inappropriate because the
dynamics is sensitive to small changes in the initial conditions.
Coarse-graining of relevant DOF is appropriate, however. The
resulting coarse-grained system will distinguish between
trajectories that circle the first or second attractor, but will
be insensitive to the details of those trajectories. In a sense,
this is analogous to the subtleties encountered in constructing
renormalization group transformations for the critical behavior of
antiferromagnets\cite{nigelbookp268,leeuwen75}.

\section{Results of coarse-graining one dimensional CA}
\label{cg_results} \subsection{Overview} The coarse-graining
procedure we described above is not constructive, but instead is a
self-consistency condition on a putative coarse-graining rule with
a specific supercell size $N$ and projection operator $P$. In many
cases the single-valuedness condition Eq.\
(\ref{trans_func_projection}) is not satisfied, the
coarse-graining fails and one must try other choices of $N$ and
$P$. It is therefore natural to ask the following questions. Can
all CA be coarse-grained? If not, which CA can be coarse-grained
and which cannot? What types of coarse-graining transitions can we
hope to find?

To answer these questions we tried systematically to coarse-grain one
dimensional CA. We considered Wolfram's 256 elementary rules and
several non-binary CA of interest to us. Our coarse-graining procedure
was applied to each rule with different choices of $N$ and $P$. In this
way we were able to coarse-grain 240 out of the 256 elementary CA.
These 240 coarse-grained-able rules include members of all four
classes. The 16 elementary CA which we could not coarse-grain are rules
30, 45, 106, 154 and their symmetries. Rules 30, 45 and 106 belong to
class 3 while 154 is a class 2 rule. We don't know if our inability to
coarse-grain these 16 rules comes from limited computing power or from
something deeper. We suspect (and give arguments in Section
\ref{Kolmogorov_complexity}) the former.

\begin{figure*}[ht]
\epsfxsize=120mm
\begin{center}
\leavevmode \epsffile{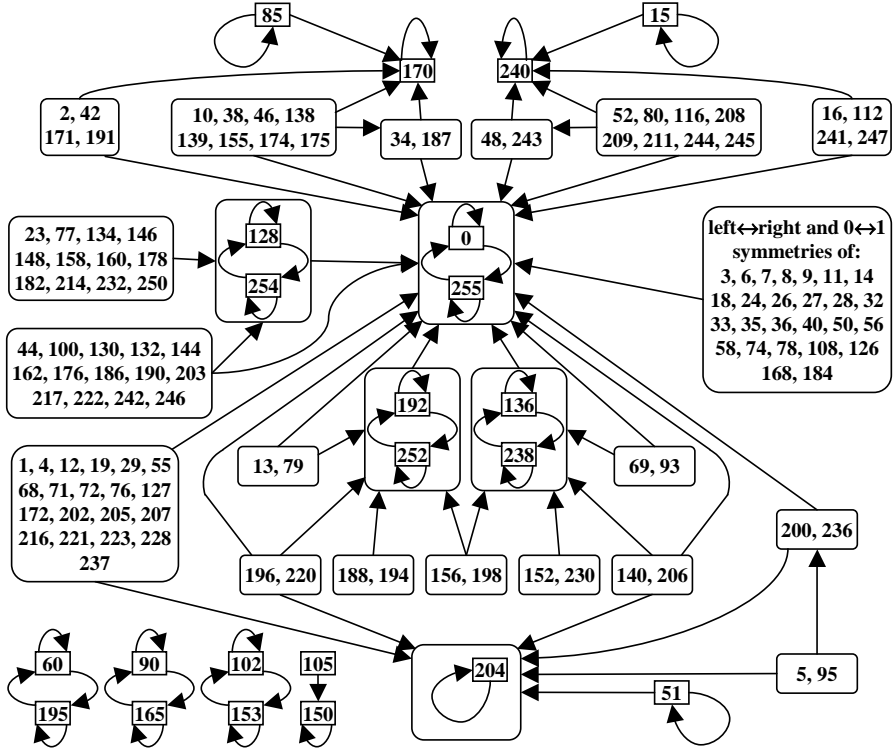}
\end{center}
\caption{Coarse-graining transitions within the
family of 256 elementary CA. Only transitions with a supercell
size $N=2,3,4$ are shown. An arrow indicates that the origin rules
can be coarse-grained by the target rules and may correspond to
several choices of $N$ and $P$. } \label{mapfigure}
\end{figure*}

The number of possible projection operators $P$ grows very fast
with $N$. Even for small $N$, it is computationally impossible
to scan all possible $P$. In order to find valid projections, we
therefore used two simple search strategies. In the first strategy,
we looked for coarse-graining transitions within the elementary CA
family by considering $P$ which project back on the binary
alphabet. Excluding the trivial projections $P(x)=0,\;\forall x$
and $P(x)=1,\;\forall x$ there are $2^{2^N}-2$ such projections.
We were able to scan all of them for $N\leq4$ and found many
coarse-graining transitions. Figure \ref{mapfigure} shows a map of
the coarse-graining transitions that we found within the family of
elementary rules. An arrow in the map indicates that each rule
from the origin group can be coarse-grained to each rule from the
target group. The supercell size $N$ and the projection $P$ are
not shown and each arrow may correspond to several choices of $N$
and $P$. As we explained above, only coarse-grainings with $N\leq
4$ are shown due to limited computing power. Other transitions
within the elementary rule family may exist with larger values of
$N$. This map is in some sense an analogue of the familiar
renormalization group flow diagrams from statistical mechanics.

Several features of Fig.\ \ref{mapfigure} are worthy of a short
discussion. First, notice that the map manifests the \lq\lq
left"$\leftrightarrow$\lq\lq right" and \lq\lq
0"$\leftrightarrow$\lq\lq 1" symmetries of the elementary CA
family. For example rules 252, 136 and 238 are the \lq\lq
0"$\leftrightarrow$\lq\lq 1", \lq\lq left"$\leftrightarrow$\lq\lq
right" and the \lq\lq 0"$\leftrightarrow$\lq\lq 1" and \lq\lq
left"$\leftrightarrow$\lq\lq right" symmetries of rule 192
respectively. Second, coarse-graining transitions are obviously
transitive, i.e. if $A$ goes to $B$ with $N_1$ and $B$ goes to $C$
with $N_2$ then $A$ goes to $C$ with $N\leq N_1\cdot N_2$. For
some transitions, the map in Fig.\ \ref{mapfigure} fails to show
this property because we did not attain large enough values of $N$.

Another interesting feature of the transition map is that the apparent
rule complexity never increases with a coarse-graining transition.
Namely, we never find a simple behaving rule which after being
coarse-grained becomes a complex rule. The transition map, therefore,
introduces a hierarchy of elementary rules and this hierarchy agrees
well with the apparent rule complexity. The hierarchy is partial and we
cannot relate rules which are not connected by a coarse-graining
transition. As opposed to the Wolfram classification, this
coarse-graining hierarchy is well defined and is therefore a good
candidate for a complexity
measure\cite{gutowitz88,gutowitz90,binder91,dubacq01,stuner90,culik88,braga95,jin03,langton90,wolf3}.

Finally notice that the eight rules 0, 60, 90, 102, 150, 170, 204,
240, whose update function has the additive form
\begin{eqnarray}
f_{\alpha\beta\gamma}\left[x_{n-1},x_n,x_{n+1}\right]&=&\alpha\cdot
x_{n-1}\oplus\beta\cdot x_n \oplus \gamma\cdot x_{n+1}\;,\nonumber \\
&&\alpha,\beta,\gamma\in \{0,1\}\;,
\end{eqnarray}
where $\oplus$ denotes the XOR operation, are all fixed points of the map. This result is not limited to
elementary rules. As showed by Barbe et.al
\cite{barbe95,barbe96,barbe97}, additive CA in arbitrary dimension
whose alphabet sizes are prime numbers coarse-grain themselves. We
conjecture that there are situations where reducible fixed points
exist for a wide range of systems, analogous to the emergence of
amplitude equations in the vicinity of bifurcation points.

When projecting back on the binary alphabet, one maximizes the
amount of information lost in the coarse-graining transition. At
first glance, this seems to be an unlikely strategy, because it is
difficult for the coarse-grained CA to emulate the original one
when so much information was lost. In terms of our coarse-graining
procedure such a projection maximizes the number of instances
$P(x)=P(y)$ of Eq.\ (\ref{trans_func_projection}). On second
examination, however. this strategy is not that poor. The fact that
there are only two states in the coarse-grained alphabet reduces
the probability that an instance $P(x)=P(y)$ of Eq.\
(\ref{trans_func_projection}) will be violated to 1/2. The
extreme case of this argument would be a projection on a
coarse-grained alphabet with a single state. Such a trivial
projection will never violate Eq.\ (\ref{trans_func_projection})
(but will never show any patterns or dynamics either).

A second search strategy for valid projection operators that we
used is located on the other extreme of the above tradeoff.
Namely, we attempt to lose the smallest possible amount of
information. We start by choosing two supercell states $z_1$ and
$z_2$ and unite them using
\begin{equation}
P_0\left(x\right)=\left\{
\begin{array}{l}
x\;,\;x\neq z_2 \\
z_1\;,\;x=z_2 \\
\end{array}
\right.\;,
\end{equation}
where the subscript in $P_0$ denotes that this is an initial trial
projection to be refined later. The refinement process of the
projection operator proceeds as follows. If $P_n$ (starting with
$n=0$) satisfies Eq.\ (\ref{trans_func_projection}) then we are
done. If on the other hand, Eq.\ (\ref{trans_func_projection}) is
violated by some
\begin{eqnarray}
P_n\left(f_{A^N}\left[x_1,x_2,x_3\right]\right)&\neq&
P_n\left(f_{A^N}\left[y_1,y_2,y_3\right]\right)\;, \nonumber \\
&&P_n(x_i)=P_n(y_i)\;,
\end{eqnarray}
the inequality is resolved by refining $P_n$ to
\begin{eqnarray}
&P_{n+1}&\left(x\right)=\left\{
\begin{array}{l}
P_n\left(x\right)\;,\;x\neq r_2\\
P_n\left(r_1\right)\;,\;x=r_2\\
\end{array}
\right.\;, \nonumber \\
r_1&=&f_{A^N}\left[x_1,x_2,x_3\right]\;\;,
r_2=f_{A^N}\left[y_1,y_2,y_3\right]\;.
\end{eqnarray}
This process is repeated until Eq.\ (\ref{trans_func_projection})
is satisfied. A non-trivial coarse-graining is found in cases
where the resulting projection operator is non-constant (more than
a single state in the coarse-grained CA).

By trying all possible $z_1,z_2$ initial pairs, the above projection
search method is guaranteed to find a valid projection if such a
projection exist on the scale defined by the supercell size $N$.
Using this method we were able to coarse-grain many CA. The
resulting coarse-grained CA that are generated in this way are
often multicolored and do not belong to the elementary CA family.
For this reason it is difficult to graphically summarize all the
transitions that we found in a map. Instead of trying to give an
overall view of those transitions we will concentrate our
attention on several interesting cases which we include in the
examples section bellow.

\subsection{Examples}
\subsubsection{Rule 105}
As our first example we choose a transition between two class 2 rules. The elementary
rule 105 is defined on the alphabet $\{\square,\blacksquare\}$
with the transition function
\begin{equation}
f_{105}\left[x_{n-1},x_n,x_{n+1}\right]=\overline{x_{n-1}\oplus
x_{n} \oplus x_{n+1}}\;,
\end{equation}
where the over-bar denotes the NOT operation, and $\square=0$, $\blacksquare=1$. We use a supercell size $N=2$ and calculate the transition
function $f_{105}^2$, defined on the alphabet
$\{\square\square,\square\blacksquare,\blacksquare\square,\blacksquare\blacksquare\}$.
Now we project this alphabet back on the
$\{\square,\blacksquare\}$ alphabet with
\begin{equation}
P\left(x\right)=\left\{\begin{array}{ll}\square, & x=\square\blacksquare,\blacksquare\square \\
\blacksquare, & x=\square\square,\blacksquare\blacksquare
\end{array}\right.\;.
\label{rule105projection}
\end{equation}
A pair of cells in rule 105 are coarse-grained to a single cell
and the value of the coarse cell is black only when the pair share
a same value. Using the above projection operator we construct the
transition function of the coarse CA. The result is found to be
the transition function of the additive rule 150:
\begin{equation}
f_{150}\left[x_{n-1},x_n,x_{n+1}\right]=x_{n-1}\oplus x_{n} \oplus
x_{n+1}\;.
\end{equation}

\begin{figure*}[t]
\centerline{ \epsfysize=55mm \epsffile{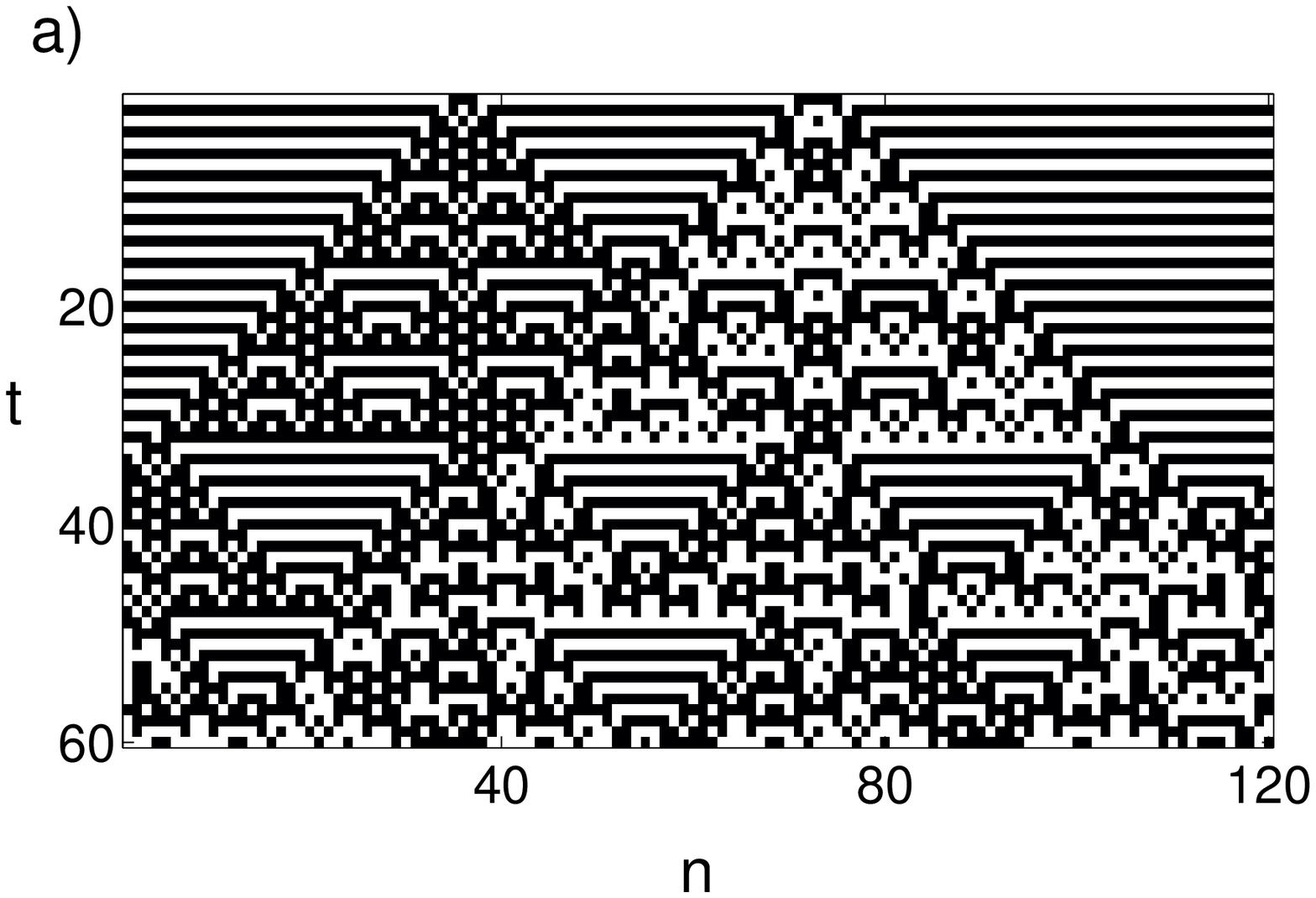}
\epsfysize=55mm \epsffile{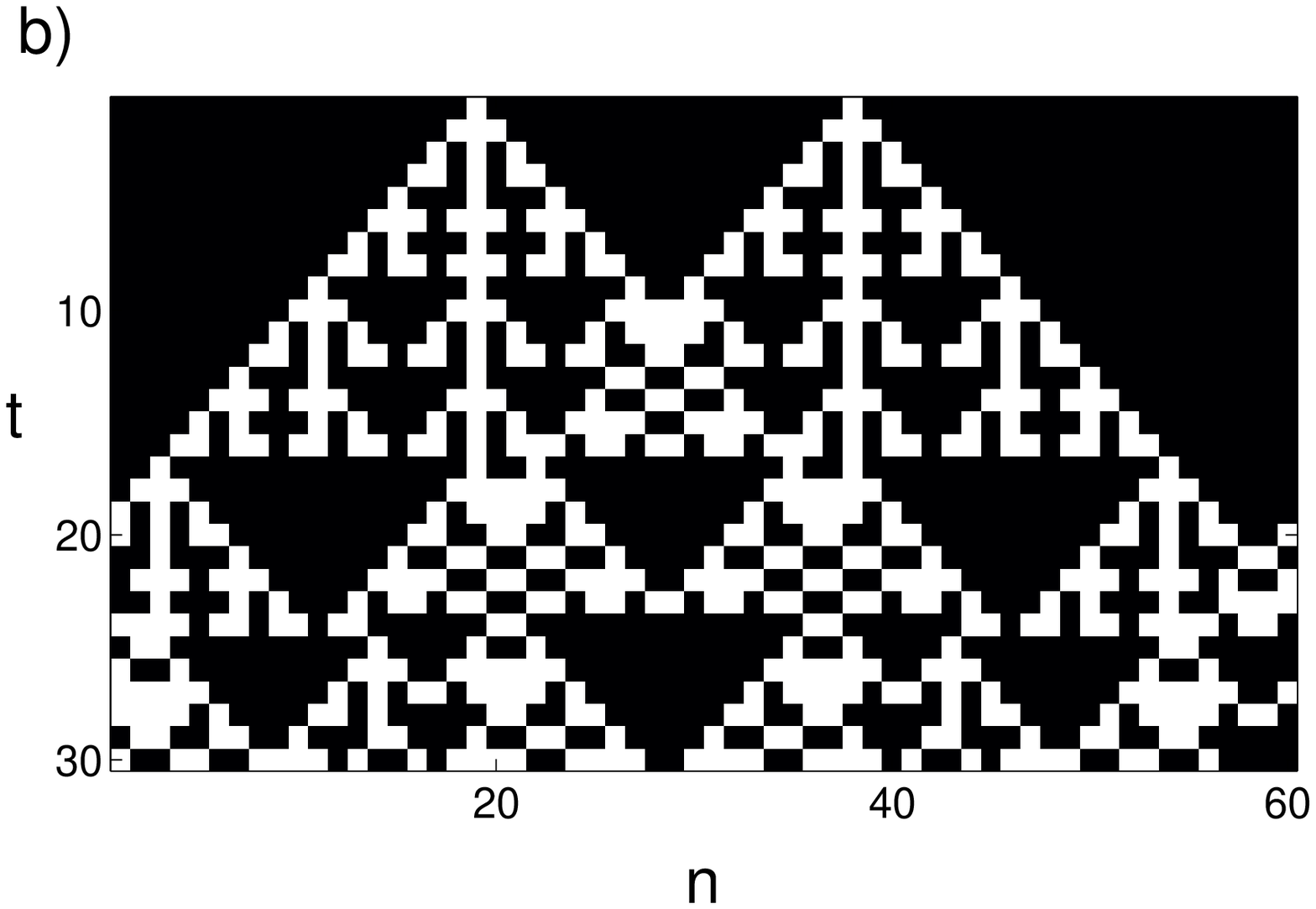}} \caption{Coarse-graining
of rule 105 by rule 150. (a) shows results of running rule 105.
The top line is the initial condition and time progress from top
to bottom. (b) shows the results of running rule 150 with the
coarse grained initial condition from (a).} \label{105to150figure}
\end{figure*}
Figure \ref{105to150figure} shows the results of this
coarse-graining transition. In Fig. \ref{105to150figure} (a) we
show the evolution of rule 105 with a specific initial condition
while Fig. \ref{105to150figure} (b) shows the evolution of rule
150 from the coarse-grained initial condition. The small scale
details in rule 105 are lost in the transformation but extended
white and black regions are coarse-grained to black regions in
rule 150. The time evolution of rule 150 captures the overall
shape of these large structures but without the black-white
decorations. As shown in Fig.\ \ref{mapfigure}, rule 150 is a
fixed point of the transition map. Rule 105 can therefore be
further coarse-grained to arbitrary scales.

\subsubsection{Rule 146}
\begin{figure*}[ht]
\centerline{ \epsfxsize=75mm \epsffile{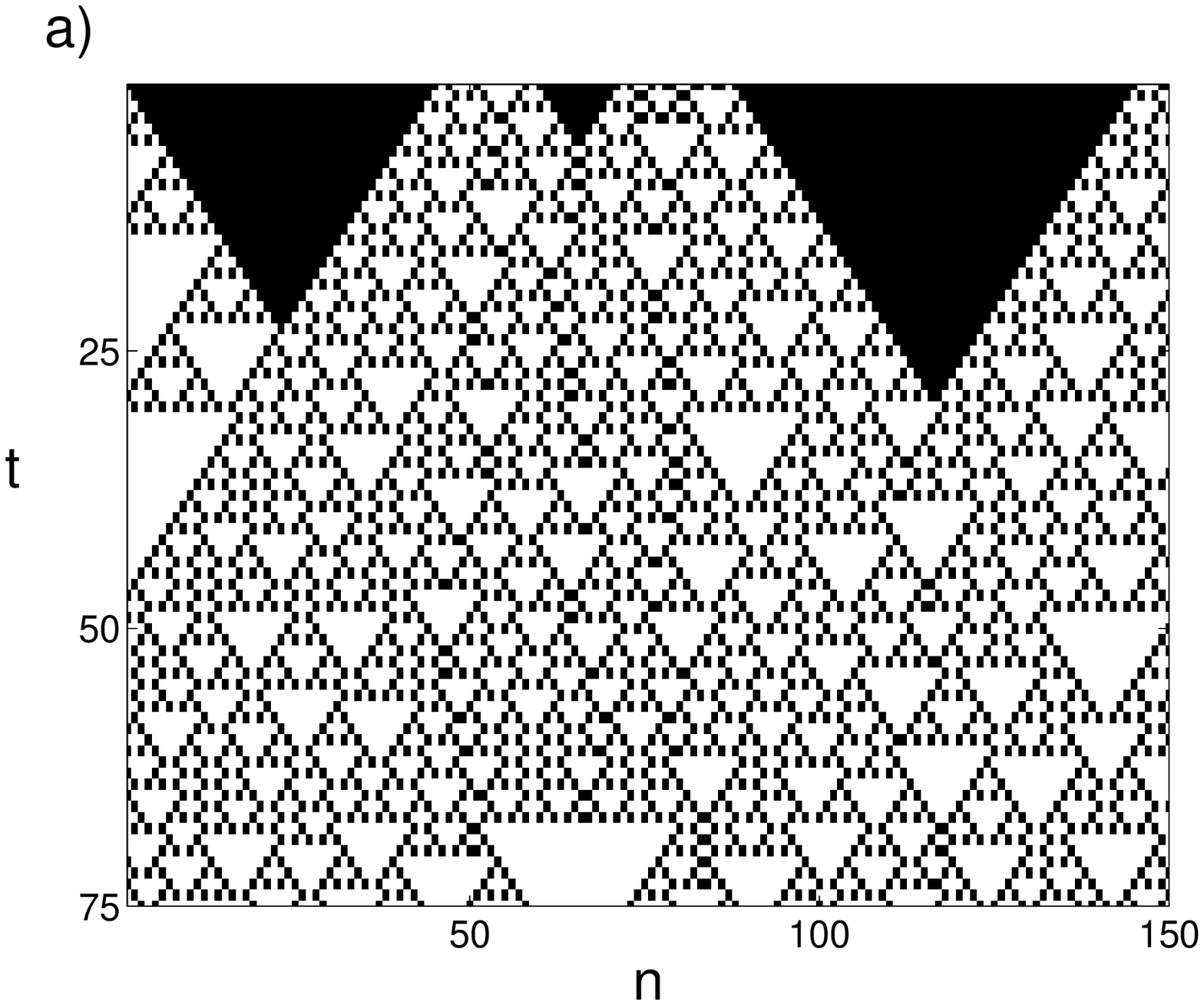}
\epsfxsize=75mm \epsffile{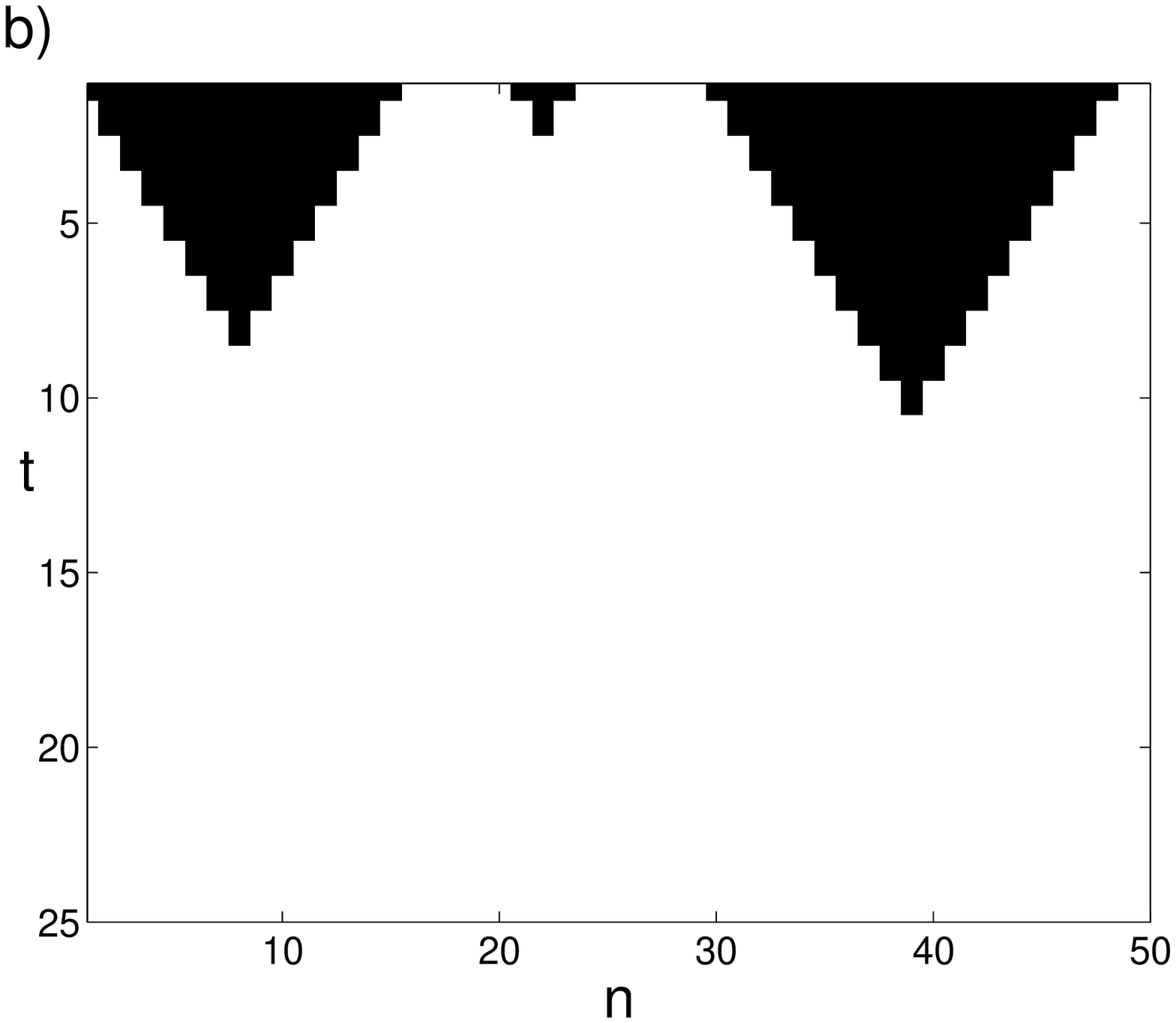}} \centerline{
\epsfxsize=75mm \epsffile{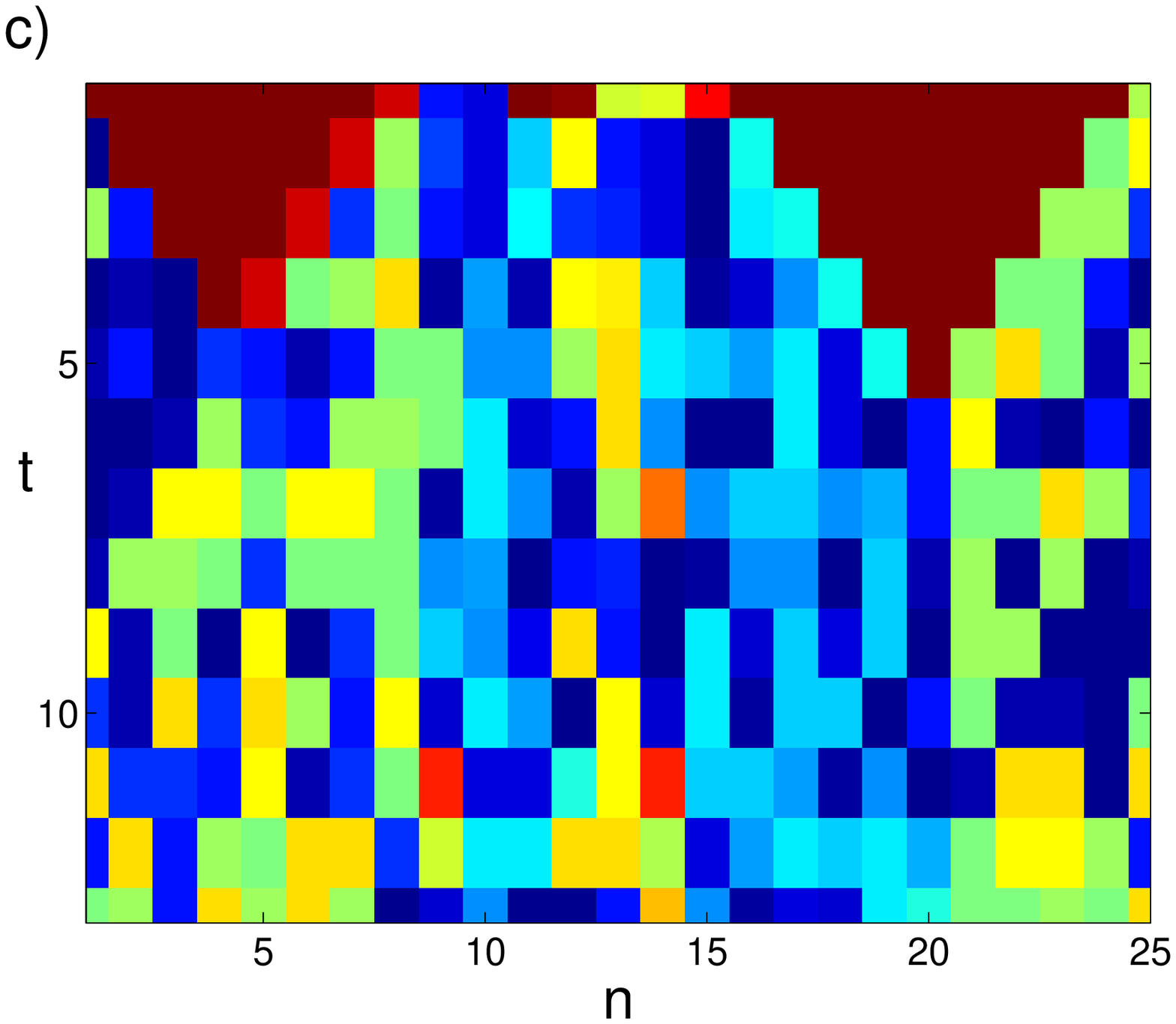} \epsfxsize=75mm
\epsffile{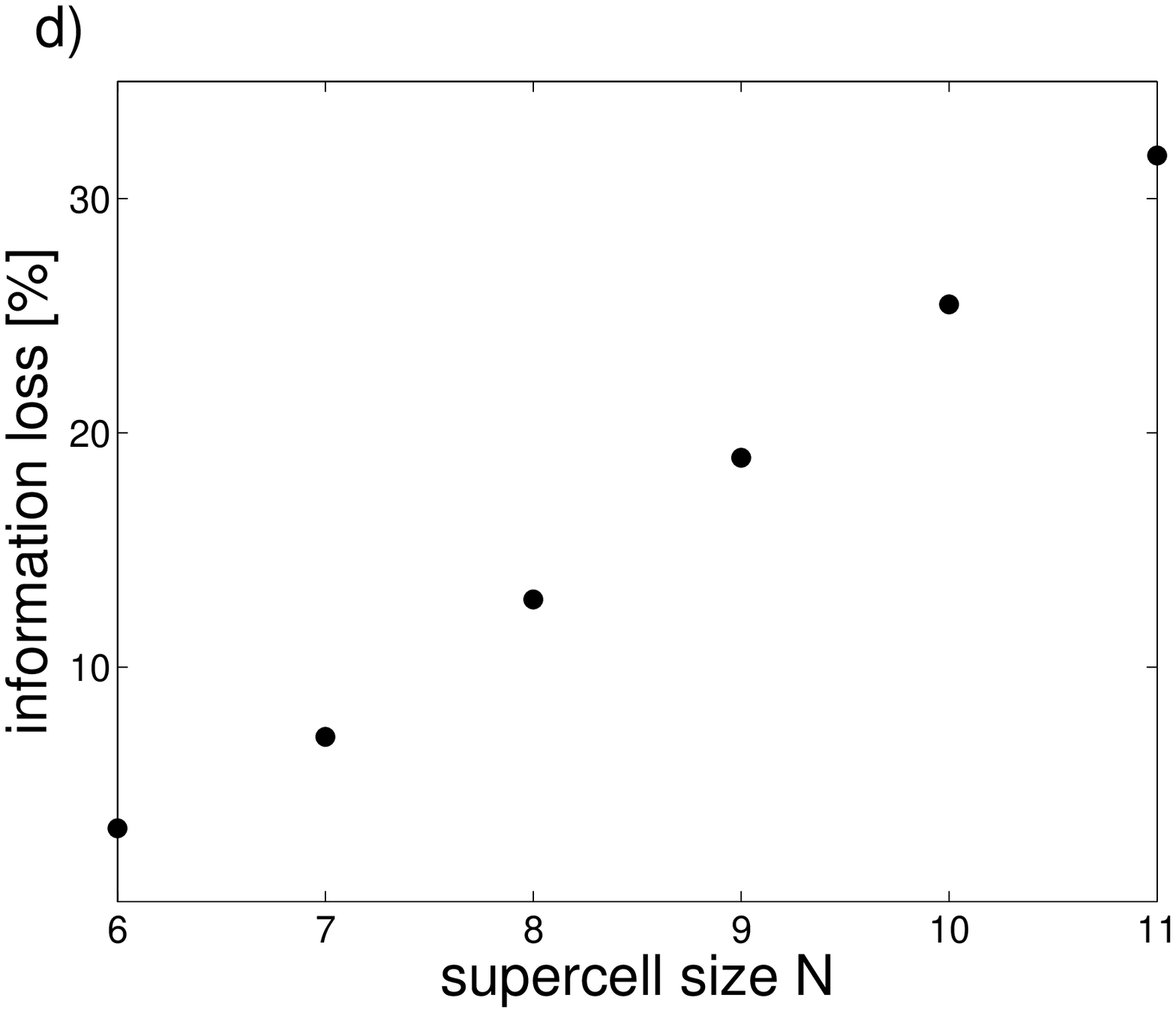}} \caption{(Color online) Coarse-graining of rule
146 by rule 128 and by a 62 color CA. (a) shows results of running
rule 146. The top line is the initial condition and time progress
from top to bottom. (b) shows the results of running rule 128 with
the coarse grained initial condition from (a). (c) shows results
of running the 62 color CA which is a coarse-grained version of
rule 146. (d) shows the percentage of supercell states that can be
eliminated when coarse graining rule 146 with different supercell
sizes $N$.} \label{cgof146figure}
\end{figure*}

As a second example of coarse-grained-able elementary CA we choose
rule 146. Rule 146 is defined on the $\{\square,\blacksquare\}$
alphabet with the transition function
\begin{eqnarray}
\lefteqn{f_{146}\left[x_{n-1},x_n,x_{n+1}\right]=} \nonumber \\
&&\left\{
\begin{array}{l}
\blacksquare\;,\;x_{n-1}x_nx_{n+1}=\square\square\blacksquare;\blacksquare\square\square;\blacksquare\blacksquare\blacksquare
\\
\square\;,\;\mbox{all other combinations} \\
\end{array}\right.
\;.
\end{eqnarray}
It produces a complex, seemingly random behavior which falls into
the class 3 group. We choose a supercell size $N=3$ and calculate the
transition function $f_{146}^3$, defined on the alphabet
$\{\square\square\square,\square\square\blacksquare, \dots
\blacksquare\blacksquare\blacksquare\}$. Now we project this
alphabet back on the $\{\square,\blacksquare\}$ alphabet with
\begin{equation}
P\left(x\right)=\left\{\begin{array}{ll}\square, &
x\neq\blacksquare\blacksquare\blacksquare \\ \blacksquare, &
x=\blacksquare\blacksquare\blacksquare
\end{array}\right.\;.
\label{rule146projection}
\end{equation}
Namely, a triplet of cells in rule 146 are coarse-grained to a
single cell and the value of the coarse cell is black only when
the triplet is all black. Using the above projection operator we
construct the transition function of the coarse CA. The result is
found to be the transition function of rule 128 which was given in
Eq.\ (\ref{f128}).
Rule 146 can therefore be coarse-grained by rule 128, a class 1
elementary CA. In Figure \ref{cgof146figure} we show the results
of this coarse-graining. Fig. \ref{cgof146figure} (a) shows the
evolution of rule 146 with a specific initial condition while Fig.
\ref{cgof146figure} (b) shows the evolution of rule 128 from the
coarse-grained initial condition. Our choice of coarse-graining
has eliminated the small scale details of rule 146. Only
structures of lateral size of three or more cells are accounted
for. The decay of such structures in rule 146 is accurately
described by rule 128.

Note that a class 3 CA was coarse-grained to a class 1 CA in the
above example. Our gain was therefore two-fold. In addition to the
phase space reduction associated with coarse-graining we have also
achieved a reduction in complexity. Our procedure was able to find
predictable coarse-grained aspects of the dynamics even though the
small scale behavior of rule 146 is complex, potentially
irreducible.

Rule 146 can also be coarse-grained by non elementary CA. Using a
supercell size of $N=6$ we found that the difference between the
combinations
$\square\blacksquare\blacksquare\blacksquare\blacksquare\square$
and $\square\blacksquare\square\square\blacksquare\square$ is
irrelevant to the long time behavior of rule 146. It is therefore
possible to project these two combinations into a single coarse
grained state. The same is true for the combinations
$\square\blacksquare\blacksquare\square\blacksquare\square$ and
$\square\blacksquare\square\blacksquare\blacksquare\square$ which
can be projected to another coarse-grained state. The end result
of this coarse-graining (Fig.\ \ref{cgof146figure} (c)) is a 62
color CA which retains the information of all other 6 cell
combinations. The amount of information lost in this transition is
relatively small, 2/64 of the supercell states have been
eliminated. More impressive alphabet reductions can be found by
going to larger scales. For $N$=7,8,9,10 and 11 we found an
alphabet reduction of 9/128, 33/256, 97/512, 261/1024 and 652/2048
respectively.  Fig.\ \ref{cgof146figure} (d) shows the percentage
of states that can be eliminated as a function of the supercell
size $N$. All of the information lost in those coarse-grainings
corresponds to irrelevant DOF.

\begin{figure}[h]
\centerline{ \epsfxsize=75mm \epsffile{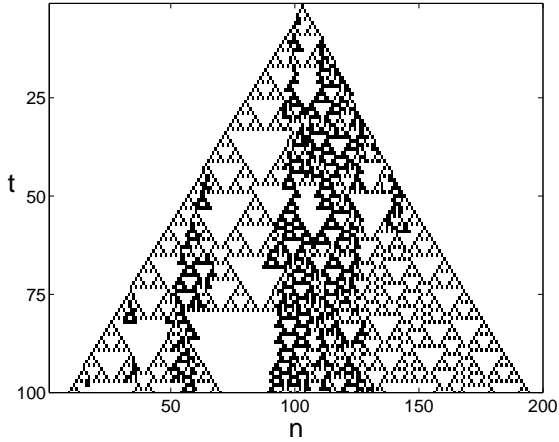}}
\caption{The sensitivity of rule 146 to a relevant DOF change in
its initial condition. The figure shows the difference (modulo 2) in the
trajectories resulting from replacing a $\square\blacksquare\square$ segment in the initial condition with $\square\blacksquare\blacksquare$.}
\label{rule146rdoffigure}
\end{figure}

The two different coarse-graining transitions of rule 146 that we
presented above are a good opportunity to show the difference
between relevant and irrelevant DOF. As we explained earlier, a
transition like 146$\rightarrow$128 where the rules has different
complexities must involve the elimination of relevant DOF. Indeed
if we modify an initial condition of rule 146 by replacing 
a $\square\blacksquare\square$ segment with
$\square\blacksquare\blacksquare$ we will get a modified
evolution. As we show in Figure \ref{rule146rdoffigure}, the
difference in the trajectories has a complex behavior and is
unbounded in space and time. However, since
$\square\blacksquare\square$ and $\square\blacksquare\blacksquare$
are both projected by Eq.\ (\ref{rule146projection}) to $\square$,
the projections of the original and modified trajectories will be identical. In contrast, the coarse graining of
rule 146 to the 62 state CA of Fig.\ \ref{cgof146figure} (c)
involves the elimination of irrelevant DOF only. If we replace a
$\square\blacksquare\blacksquare\square\blacksquare\square$ in the
initial condition with a
$\square\blacksquare\square\blacksquare\blacksquare\square$ we
find that the difference between the modified and unmodified
trajectories decays after a few time steps.

\subsubsection{Rule 184}
The elementary CA rule 184 is a simplified one lane traffic flow
model. Its transition function is given by
\begin{eqnarray}
\lefteqn{f_{184}\left[x_{n-1},x_n,x_{n+1}\right]=} \nonumber \\
&&\left\{
\begin{array}{l}
\square\;,\;x_{n-1}x_nx_{n+1}=\square\square\square;\square\square\blacksquare;\square\blacksquare\square;\blacksquare\blacksquare\square
\\
\blacksquare\;,\;x_{n-1}x_nx_{n+1}=\square\blacksquare\blacksquare;\blacksquare\square\square;\blacksquare\square\blacksquare;\blacksquare\blacksquare\blacksquare \\
\end{array}\right.
\;.
\end{eqnarray}
Identifying a black cell with a car moving to the right and a
white cell with an empty road segment we can rewrite the update
rule as follows. A car with an empty road segment to its right
advances and occupies the empty segment. A car with another car to
its right will avoid a collision and stay put. This is a
deterministic and simplified version of the more realistic Nagel
Schreckenberg model \cite{nagel_schreckenberg}.

Rule 184 can be coarse-grained to a 3 color CA using a supercell
size $N=2$ and the local density projection
\begin{equation}
P\left(x\right)=\left\{
\begin{array}{l}
\square\;,\;x=\square\square \\
\greysquare\;,\;x=\square\blacksquare;\blacksquare\square \\
\blacksquare\;,\;x=\blacksquare\blacksquare \\
\end{array}\right.\;.
\end{equation}
The update function of the resulting CA is given by
\begin{eqnarray}
\lefteqn{f\left[y_{n-1},y_n,y_{n+1}\right]=} \\
&&\left\{
\begin{array}{l}
\square\;,\;y_{n-1}y_ny_{n+1}=\square\square\square;\square\square\greysquare;\square\square\blacksquare;\square\greysquare\square;\square\greysquare\greysquare \\
\blacksquare\;,\;y_{n-1}y_ny_{n+1}=\square\blacksquare\blacksquare;\greysquare\greysquare\blacksquare;\greysquare\blacksquare\blacksquare;\blacksquare\greysquare\blacksquare;\blacksquare\blacksquare\blacksquare \\
\greysquare\;,\;\mbox{all other combinations} \\
\end{array}\right.. \nonumber
\end{eqnarray}

\begin{figure*}[tb]
\centerline{ \epsfxsize=75mm \epsffile{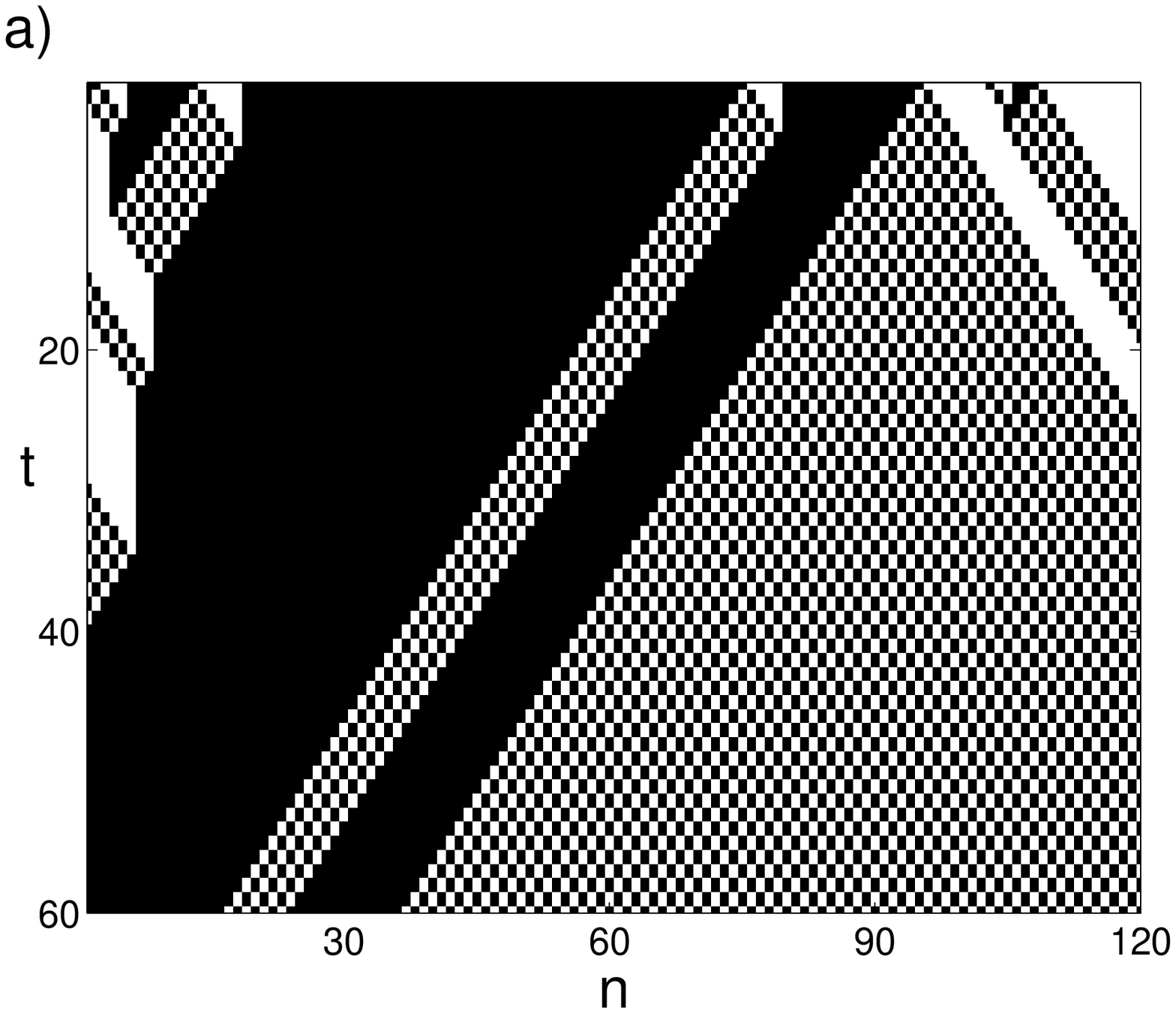}
\epsfxsize=75mm \epsffile{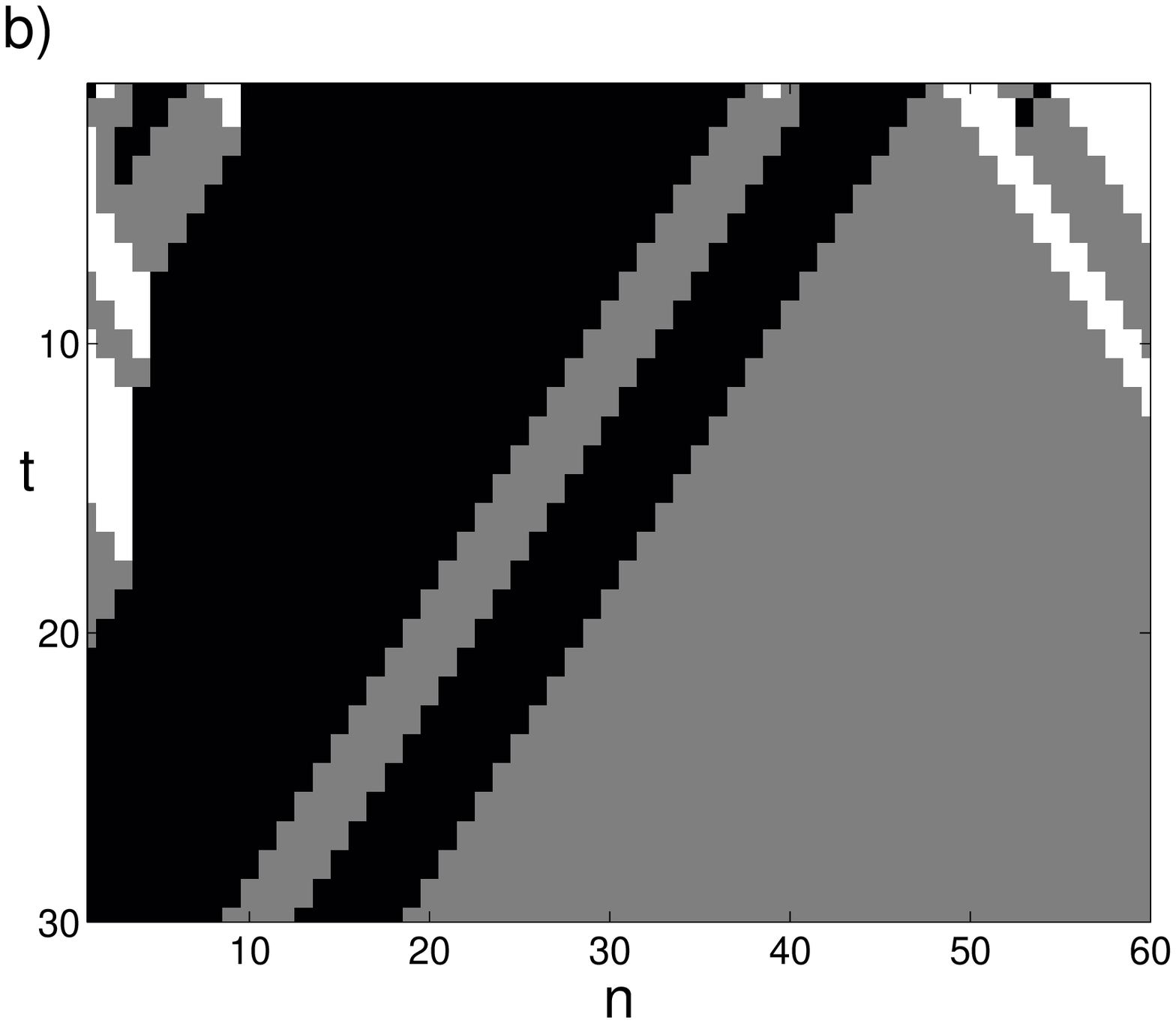}} \centerline{ \epsfxsize=75mm
\epsffile{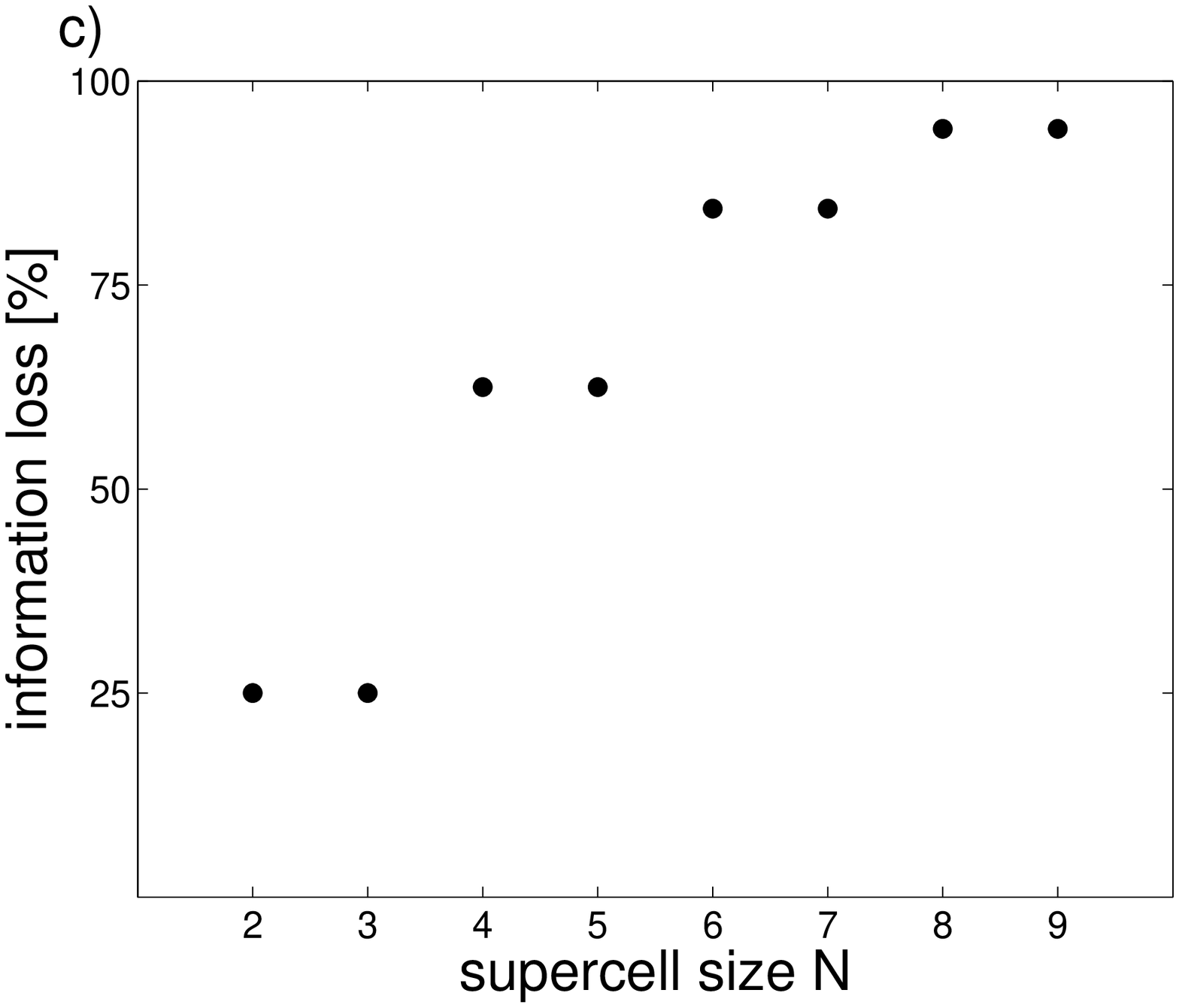}} \caption{Coarse graining of rule
184 by a 3 state CA. (a) shows a trajectory of rule 184. (b) shows
the corresponding trajectory of the coarse-grained CA. (c) shows
the percentage of supercell states that can be eliminated when
coarse graining rule 184 with different supercell sizes $N$.}
\label{rule184figure}
\end{figure*}

Figure \ref{rule184figure} shows the result of this
coarse-graining. Fig.\ \ref{rule184figure} (a) shows a trajectory
of rule 184 while Fig.\ \ref{rule184figure} (b) shows the
trajectory of the coarse CA. From this figure it is clear that the
white zero density regions correspond to empty road and the black
high density regions correspond to traffic jams. The density 1/2
grey regions correspond to free flowing traffic with an exception
near traffic jams due to a boundary effect.

By using larger supercell sizes it is possible to find other
coarse-grained versions of rule 184. As in the above example, the
coarse-grained states group together local configurations of equal
car densities. The projection operators however are not functions
of the local density alone. They are a partition of such a
function and there could be several coarse-grained states which
correspond to the same local car density. We found (empirically)
that for even supercell sizes $N=2k$ the coarse-grained CA contain
$k^2/2+3k/2+1$ states and for odd supercell sizes $N=2k+1$ they
contain $k^2+3k+2$ states. Figure \ref{rule184figure} (c) shows
the amount of information lost in those transitions as a function
of $N$. Most of the lost information corresponds to relevant DOF
but some of it is irrelevant.

\subsubsection{Rule 110}
Rule 110 is one of the most interesting rules in the elementary CA
family. It belongs to class 4 and exhibits a complex behavior
where several types of \lq\lq particles" move and interact above a
regular background. The behavior of these \lq\lq particles" is
rich enough to support universal computation \cite{nks}. In this
sense rule 110 is maximally complex because it is capable of
emulating all computations done by other computing devices in
general and CA in particular. As a consequence it is also
undecidable \cite{wolf2}.

\begin{figure*}[ht]
\centerline{ \epsfxsize=75mm \epsffile{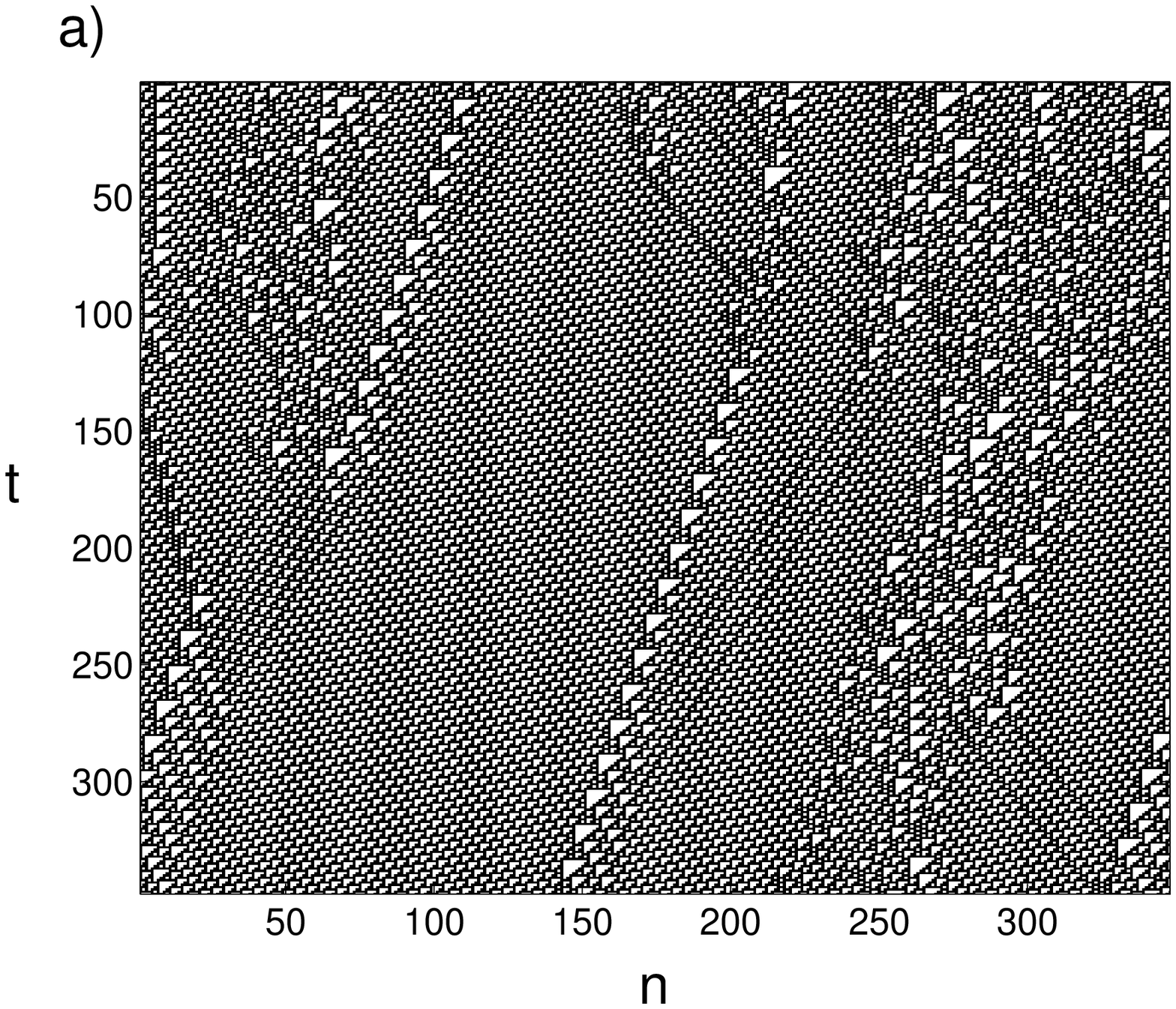}
\epsfxsize=75mm \epsffile{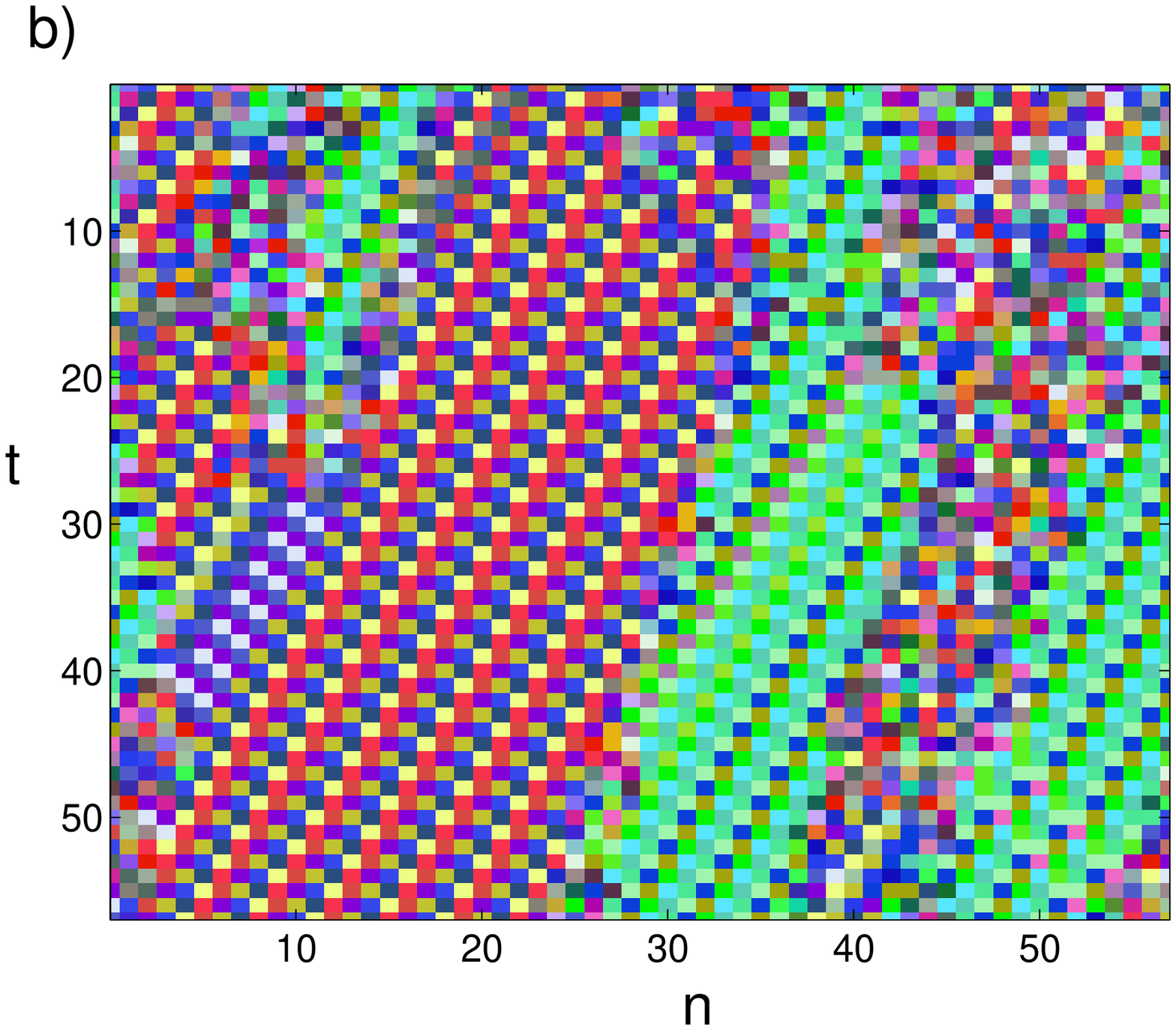}} \centerline{
\epsfxsize=75mm \epsffile{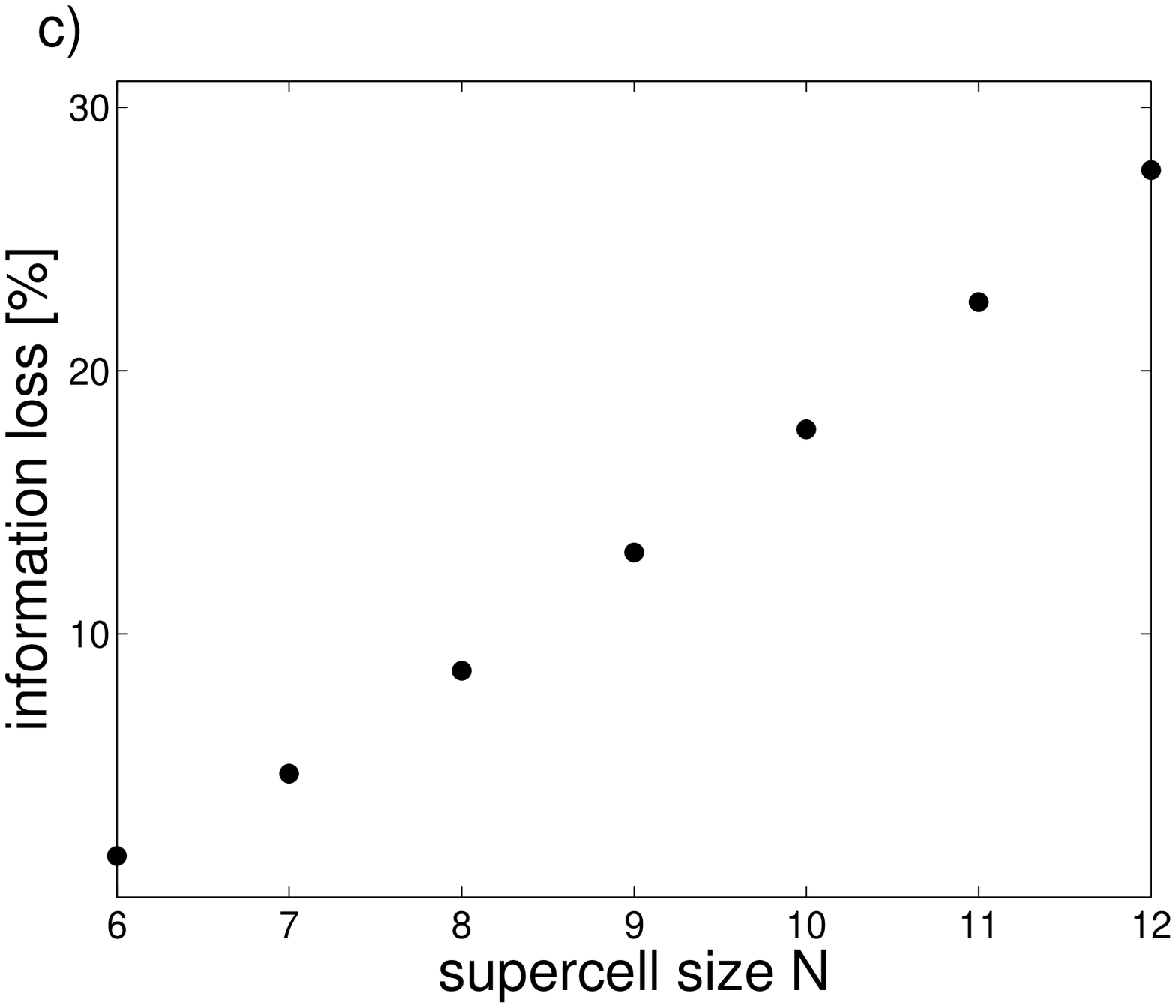} \epsfxsize=75mm
\epsffile{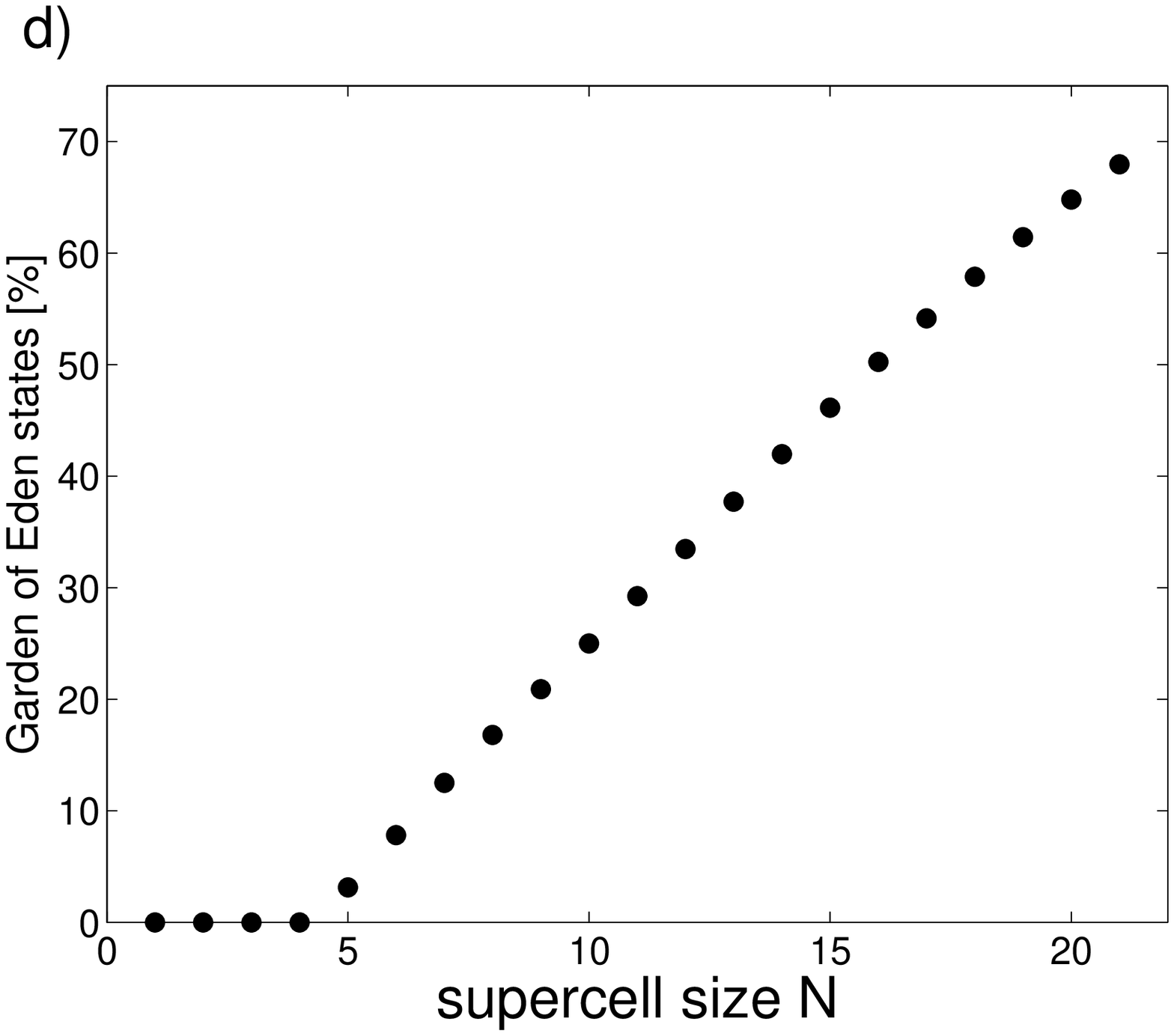}} \caption{(Color online) Coarse graining of
rule 110. (a) shows a trajectory of rule 110. (b) shows a coarse
graining of rule 110 by a 63 color CA. (c) shows the percentage of
supercell states that can be eliminated when coarse graining rule
110 with different supercell sizes $N$. (d) shows the percentage
of \lq\lq Garden of Eden" states out of the $2^N$ possible states
of supercell $N$ versions of rule 110.} \label{rule110figure}
\end{figure*}

We found several ways to coarse-grain rule 110. Using $N=6$, it is
possible to project the 64 possible supercell states onto an
alphabet of 63 symbols. Figure \ref{rule110figure} (a) and (b)
shows a trajectory of rule 110 and the corresponding trajectory of
the coarse-grained 63 states CA. A more impressive reduction in
the alphabet size is obtained by going to larger values of $N$.
For $N=7,8,9,10,11,12$ we found an alphabet reduction of $6/128$,
$22/256$, $67/512$, $182/1024$, $463/2048$ and $1131/4096$
respectively. Only irrelevant DOF are eliminated in those
transitions. Fig.\ \ref{rule110figure} (c) shows the percentage of
reduced states as a function of the supercell size $N$. We expect
this behavior to persist for larger values of $N$.

Another important coarse-graining of rule 110 that we found is the
transition to rule 0. Rule 0 has the trivial dynamics where all
initial states evolve to the null configuration in a single time
step. The transition to rule 0 is possible because many cell
sequences cannot appear in the long time trajectories of rule 110.
For example the sequence
$\square\blacksquare\square\blacksquare\square$ is a so called
\lq\lq Garden of Eden" of rule 110. It cannot be generated by rule
110 and can only appear in the initial state. Coarse-graining by
rule 0 is achieved in this case using $N=5$ and projecting
$\square\blacksquare\square\blacksquare\square$ to $\blacksquare$
and all other five cell combinations to $\square$. Another example
is the sequence
$\square\blacksquare\square\blacksquare\blacksquare\square\square\square\blacksquare\blacksquare\square\square\blacksquare$.
This sequence is a \lq\lq Garden of Eden" of the $N=13$ supercell
version of rule 110. It can appear only in the first 12 time steps
of rule 110 but no later. Coarse-graining by rule 0 is achieved in
this case using $N=13$ and projecting
$\square\blacksquare\square\blacksquare\blacksquare\square\square\square\blacksquare\blacksquare\square\square\blacksquare$
to $\blacksquare$ and all other 13 cell combinations to $\square$.
These examples are important because they show that even though
rule 110 is undecidable it has decidable and predictable
coarse-grained aspects (however trivial). To our knowledge rule
110 is the only proven undecidable elementary CA
and therefore this is the only (proven) example of undecidable to
decidable transition that we found within the elementary CA
family.

It is interesting to note that the number of \lq\lq Garden of
Eden" states in supercell versions of rule 110 grows very rapidly
with the supercell size $N$. As we show in Fig.\
\ref{rule110figure} (d), the fraction of \lq\lq Garden of Eden"
states out of the $2^N$ possible sequences, grows almost linearly
with $N$. In addition, at every scale $N$ there are new \lq\lq
Garden of Eden" sequences which do not contain any smaller
\lq\lq Gardens of Eden" as subsequences. These results are
consistent with our understanding that even though the dynamics
looks complex, more and more structure emerges as one goes to
larger scales. We will have more to say about this in section
\ref{Kolmogorov_complexity}.

The \lq\lq Garden of Eden" states of supercell versions of rule
110 represent pieces of information that can be used in reducing
the computational effort in rule 110. The reduction can be
achieved by truncating the supercell update rule to be a function
of only non \lq\lq Garden of Eden" states. The size of the
resulting rule table will be much smaller ($\approx 3\%$ with
$N=21$) than the size of the supercell rule table. Efficient
computations of rule 110 can then be carried out by running rule
110 for the first $N$ time steps. After $N$ time steps the system
contains no \lq\lq Garden of Eden" sequences and we can continue
to propagate it by using the truncated supercell rule table
without loosing any information. Note that we have not reduced
rule 110 to a decidable system. At every scale we achieved a
constant reduction in the computational effort. Wolfram
has pointed out that many irreducible systems have pockets of
reducibility and termed such a reduction as \lq \lq superficial
reducibility" (see page 746 in Ref.\ \onlinecite{nks}). It will be
interesting to check how much \lq \lq superficial reducibility" is
contained in rule 110 at larger scales. It will be inappropriate
to call it \lq\lq superficial" if the curve in Fig.\
\ref{rule110figure} (d) approaches 100\% in the large $N$ limit.

\subsubsection{Albert and Culik universal CA}
It might be argued that the coarse-graining of rule 110 by rule 0
is a trivial example of an undecidable to a decidable
coarse-graining transition. The fact that certain configurations
cannot be arrived at in the long time behavior is not very
surprising and is expected of any irreversible system. In order to
search for more interesting examples we studied other one
dimensional universal CA that we found in the literature. Lindgren
and Nordahl \cite{lindgren90} constructed a 7 state nearest neighbor and
a 4 state next-nearest neighbor CA that are capable of emulating a
universal Turing machine. The entries in the update tables of
these CA are only partly determined by the emulated Turing machine
and can be completed at will. We found that for certain completion
choices these two universal CA can be coarse-grained to a trivial
CA which like rule 0 decay to a quiescent configuration in a
single time step. Another universal CA that can undergo such a
transition is Wolfram's 19 state, next-nearest neighbor universal
CA \cite{nks}. These results are essentially equivalent to the
rule 110 $\rightarrow$ rule 0 transition.

\begin{figure*}[ht]
\centerline{ \epsfxsize=75mm \epsffile{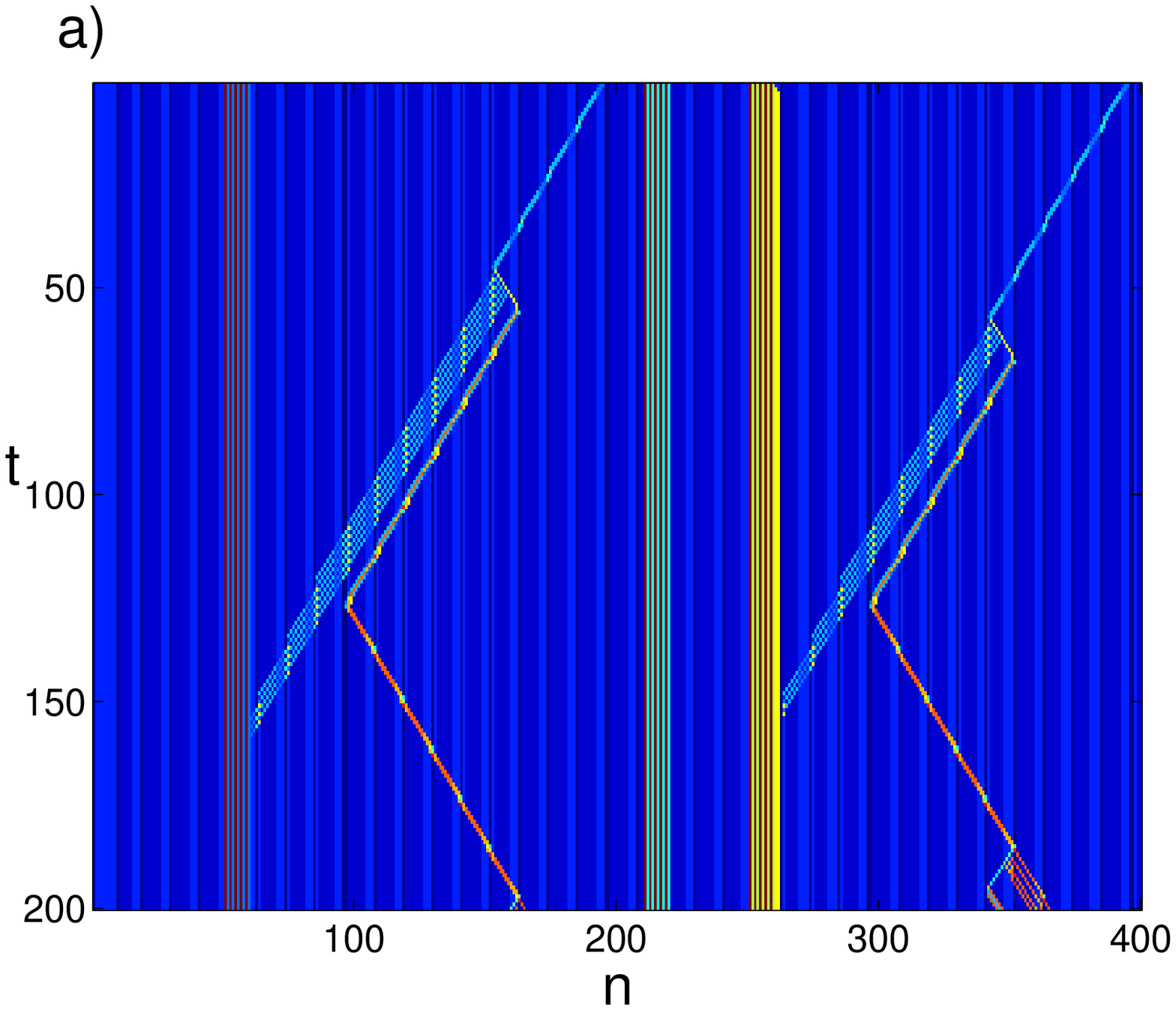}
\epsfxsize=75mm \epsffile{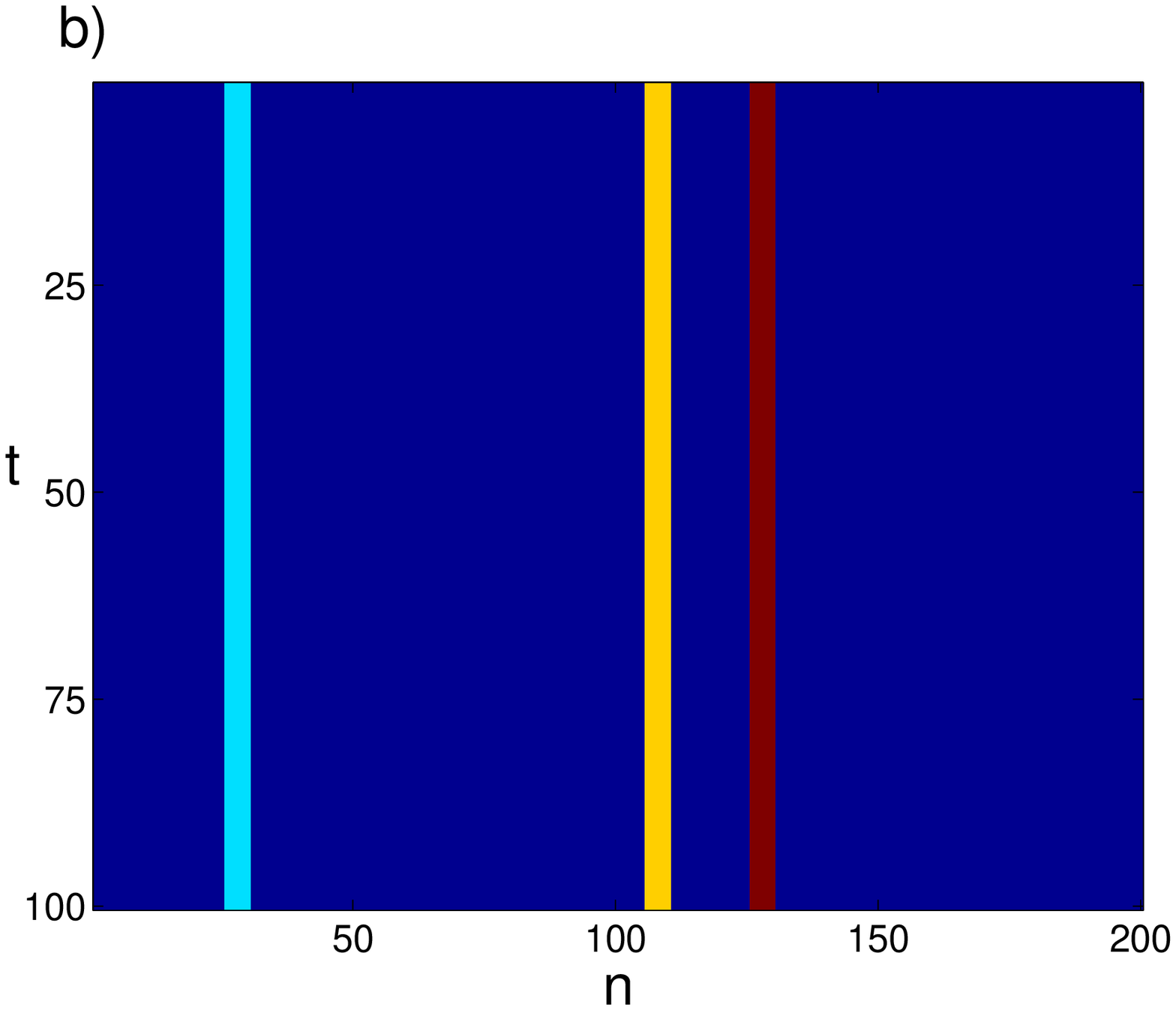}}  \caption{(Color online) Coarse
graining of Albert and Culik's \cite{albert87} 14 states universal CA
by a 4 state copy CA. (a) shows a trajectory of Albert and Culik's
universal CA while (b) shows the corresponding trajectory of the
coarse-grained CA. The supercell size used here is $N=2$}
\label{AlbertCulik_figure}
\end{figure*}

A more interesting example is Albert and Culik's \cite{albert87}
universal CA. It is a 14 state nearest-neighbor CA which is
capable of emulating all other CA. The transition table of this CA
is only partly determined by its construction and can be
completed at will. We found that when the empty entries in the
transition function are filled by the copy operation
\begin{equation}
f\left[x_{n-1},x_n,x_{n+1}\right]=x_n\;, \label{copy_operation}
\end{equation}
the resulting undecidable CA has many coarse-graining transitions
to decidable CA. In all these transitions the coarse-grained CA
performs the copy operation Eq.\ (\ref{copy_operation}) for all
$\left(x_{n-1},x_n,x_{n+1}\right)$. Different transitions differ
in the projection operator and the alphabet size of the
coarse-grained CA. Figure \ref{AlbertCulik_figure} shows a
coarse-graining of Albert and Culik's universal CA to a 4 state
copy CA. The coarse-grained CA captures three types of persistent
structures that appear in the original system but is ignorant of
more complicated details. The supercell size used here is $N=2$.

\section{Coarse-grain-ability of local processes}
\label{Kolmogorov_complexity} In the previous section we showed
that a large majority of elementary CA can be coarse-grained in
space and time. This is rather surprising since finding a valid
projection operator is equivalent to solving Eq.\
(\ref{cg_matrix_form}) which is greatly over constrained.
Solutions for this equation should be rare for random choices of
the matrix $A^N$. In this section we show that solutions of Eq.\
(\ref{cg_matrix_form}) are frequent because $A^N$ is not random
but a highly structured object. As the supercell size $N$ is
increased, $A^N$ becomes less random and the probability of finding
a valid projection approaches unity.

To appreciate the high success rate in coarse-graining elementary
CA consider the following statistics. By using supercells of size
$N=2$ and considering all possible projection operators
$P:\{0\dots 3\}\rightarrow \{0,1\}$ we were able to coarse-grain
approximately one third of all 256 elementary CA rules. Recall
that the coarse-graining procedure that we use involves two
stages. In the first stage we generate the supercell version
$A^N$, a 4 color CA in the $N=2$ case. In the second stage we look
for valid projection operators. 4 color CA that are $N=2$ supercell
versions of elementary CA are a tiny fraction of all possible
($4^{(4^3)}\approx 3\times 10^{38}$) 4 color CA. If we pick a
random 4 color CA and try to project it; i.e. attempt to solve
Eq.\ (\ref{cg_matrix_form}) with $A^N$ replaced by an arbitrary 4
color CA, we find an average of one solvable instance out of every
$\approx 1.6\times 10^7$ attempts. This large difference in the
projection probability indicates that 4 color CA which are
supercells versions of elementary rules are not random. The
numbers become more convincing when we go to larger values of $N$
and attempt to find projections to random $2^N$ color CA.

To put our arguments on a more quantitative level we need to
quantify the information content of supercell versions of CA. An
accepted measure in algorithmic information theory for the
randomness and information content of an individual\cite{entropy_comment} object is its
Kolmogorov complexity (algorithmic complexity)
\cite{kol_com_intro,chaitin87}. The Kolmogorov complexity $K_U(x)$
of a string of characters $x$ with respect to a universal computer
$U$ is defined as
\begin{equation}
K_U\left(x\right)=\frac{L_U(x)}{length(x)}\;,
\label{Kolmogorov_complexity_def}
\end{equation}
where $length(x)$ is the length of $x$ in bits and $L_U(x)$ is the
bit length of the minimal computer program that generates $x$ and
halts on $U$ (irrespective of the running time). This definition
is sensitive to the choice of machine $U$ only up to an additive
constant in $L_U(x)$ which do not depend on $x$. For long strings
this dependency is negligible and the subscript $U$ can be
dropped. According to this definition, strings which are very
structured require short generating programs and will therefore
have small Kolmogorov complexity. For example, a periodic $x$ with
period $p$ can be generated by a $\sim p$ long program and
$K(x)\sim p/length(x)$. In contrast, if $x$ has no structure it
must be generated literally, i.e. the shortest program is \lq\lq
print(x)". In such cases $L(x)\sim length(x)$, $K(x)\sim 1$ and
the information content of $x$ is maximal. By using simple
counting arguments \cite{kol_com_intro} it is easy to show that
simple objects are rare and that $K(x)\sim 1$ for most objects
$x$. Kolmogorov complexity is a powerful and elegant concept which
comes with an annoying limitation. It is uncomputable, i.e. it is
impossible to find the length of the minimal program that
generates a string $x$. It is only possible to bound it.

It is easy to see that supercell CA are highly structured objects
by looking at their Kolmogorov complexity. Consider the CA
$A=\left(a\left(t\right),S,f_A\right)$ and its $N$'th supercell
version $A^N=\left(a^N,S^N,f_{A^N}\right)$ (for simplicity of
notation we omit the subscript $A$ from the alphabet size). The
transition function $f_{A^N}$ is a table that specifies a cell's new
state for all $S^{3N}$ possible local configurations (assuming A
is nearest neighbor and one dimensional). $f_{A^N}$ can therefore
be described by a string of $S^{3N}$ symbols from the alphabet
$\{0\dots S^N-1\}$. The bit length $length\left(f_{A^N}\right)$ of
such a description is
\begin{equation}
length\left(f_{A^N}\right)=S^{3N}\cdot N\log_2{S}\;.
\end{equation}

If $A^N$ was a typical CA with $S^N$ colors we could expect that
$L\left(f_{A^N}\right)$, the length of the minimal program that
generates $f_{A^N}$, will not differ significantly from
$length\left(f_{A^N}\right)$. However, since $A^N$ is a super cell
version of $A$ we have a much shorter description, i.e. to
construct $A^N$ from $A$. This construction involves running $A$,
$N$ time steps for all possible initial configurations of $3N$
cells. It can be conveniently coded in a program as repeated
applications of the transition function $f_A$ within several
loops. Up to an additive constant\cite{kol_com_intro}, the length
of such a program will be equal to the bit length description of
$f_A$:
\begin{equation}
\tilde{L}\left(f_{A^N}\right)=S^3\cdot log_2{S}\;.
\end{equation}
Note that we have used $\tilde{L}$ to indicate that this is an
upper bound for the length of the minimal program that generates
$f_{A^N}$. This upper bound, however, should be tight for an update
rule $f_A$ with little structure. The Kolmogorov complexity of
$f_{A^N}$ can consequently be bounded by
\begin{equation}
K\left(f_{A^N}\right)\leq
\tilde{K}\left(f_{A^N}\right)=\frac{\tilde{L}\left(f_{A^N}\right)}{length\left(f_{A^N}\right)}=N^{-1}\cdot
S^{3(1-N)}\;. \label{upper_bound_kc}
\end{equation}
This complexity approaches zero at large values of $N$.

Our argument above shows that the large scale behavior of CA (or
any local process) must be simple in some sense. We would like to
continue this line of reasoning and conjecture that the small
Kolmogorov complexity of the large scale behavior is related to
our ability to coarse-grain many CA. At present we are unable to
prove this conjecture analytically, and must therefore resort to
numerical evidence which we present below.

\subsection{Garden of Eden states of supercell CA}
\label{GardensOfEden}
Ideally, in order to show that such a connection exists one would
attempt to coarse-grain CA with different alphabets and on
different length scales (supercell sizes), and verify that the
success rate correlates with the Kolmogorov complexity of the
generated supercell CA. This, however, is computationally very
challenging and going beyond CA with a binary alphabet and
supercell sizes of more than $N=4$ is not realistic. A more modest
experiment is the following. We start with a CA $A$ with an
alphabet $S$, and check whether its $N$ supercell version $A^N$
contains all possible $S^N$ states. Namely, if there exist $x\in
\{0\dots S^N-1\}$ such that
\begin{equation}
f_{A^N}(y_1,y_2,y_3)\neq x\;, \forall y_1,y_2,y_3 \in \{0\dots
S^N-1\}\;.
\end{equation}
Such a missing state of $A^N$ is sometimes referred to as a \lq\lq
Garden of Eden" configuration because it can only appear in the
initial state of $A^N$. Note that by the construction of $A^N$, a
\lq\lq Garden of Eden" state of $A^N$ can appear only in the first
$N-1$ time steps of $A$ and is therefore a generalized \lq\lq
Garden of Eden" of $A$. In cases where a state of $A^N$ is
missing, $A$ can be trivially coarse-grained to the elementary CA
rule 0 by projecting the missing state of $A^N$ to \lq\lq 1" and
all other combinations to \lq\lq 0". This type of trivial
projection was discussed earlier in connection with the
coarse-graining of rule 110. Finding a \lq\lq Garden of Eden"
state of $A^N$ is computationally relatively easy because there is
no need to calculate the supercell transition function $f_{A^N}$.
It is enough to back-trace the evolution of $A$ and check if all
$N$ cell combinations has a $3N$ cell ancestor combination, $N$
time steps in the past.

\begin{figure*}[ht]
\centerline{ \epsfxsize=75mm \epsffile{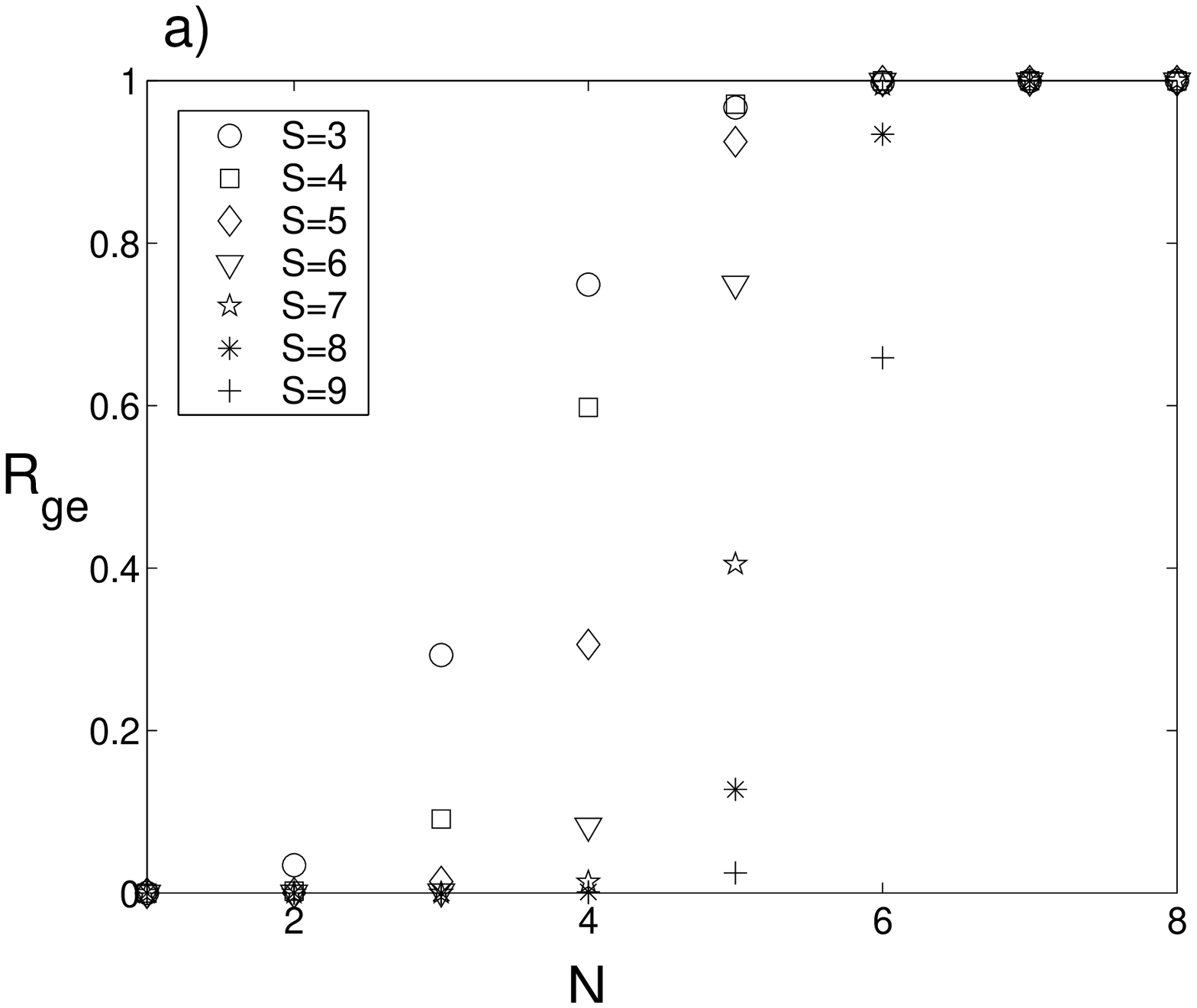}
\epsfxsize=75mm \epsffile{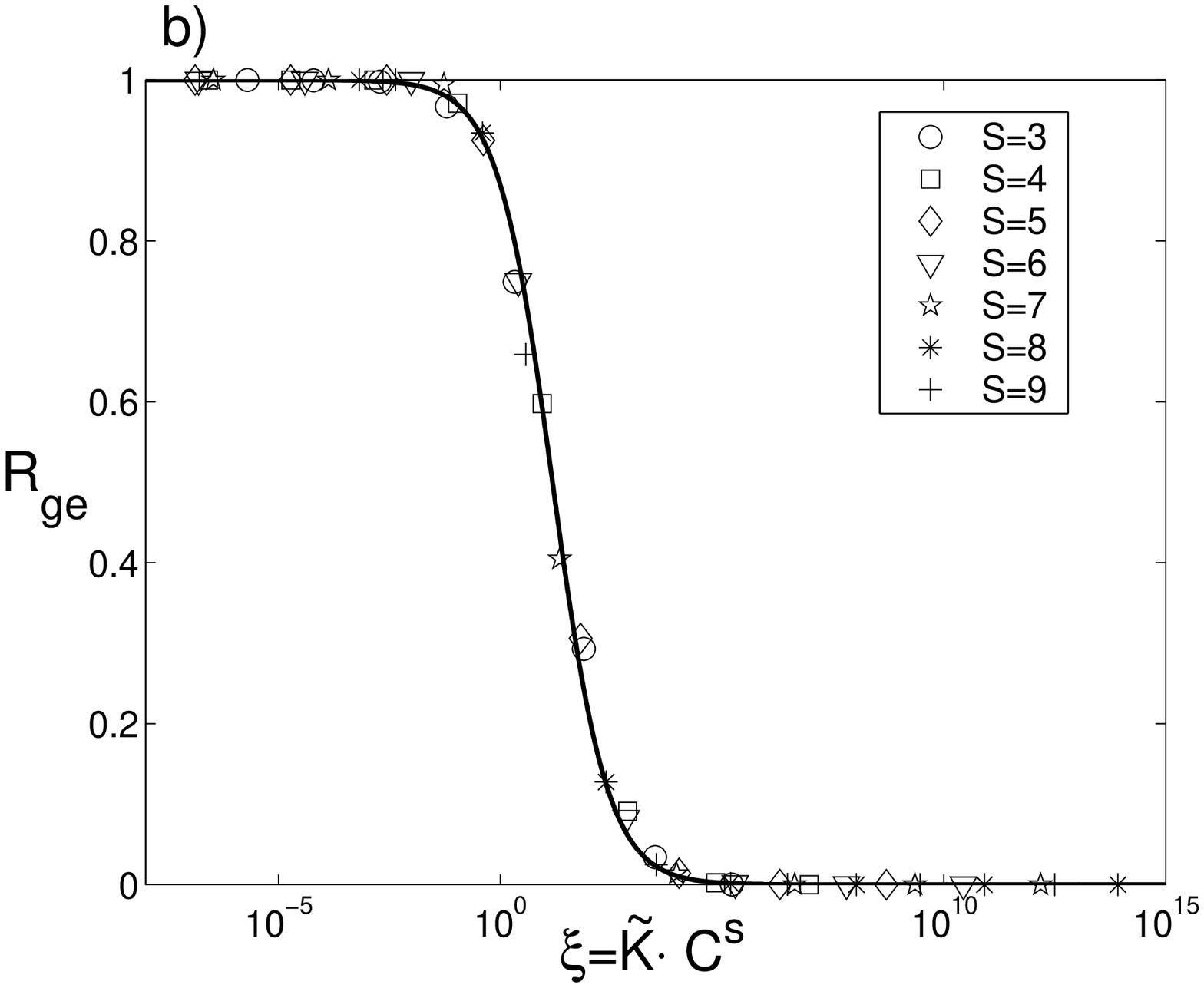}} \caption{(a)
The fraction $R_{ge}$ of CA whose $N$'th supercell version has at
least one missing state. Different symbols correspond to different
alphabet sizes $S$ of the original CA. (b) Data collapse of the
curves $R_{ge}(N,S)$ from a) when plotted against the scaling
variable $\xi=\tilde{K}(N,S)\cdot C^S$. The solid line shows that
the scaling function can be fitted by Eq.\
(\ref{missing_states_scaling_fit}).} \label{missing_colors}
\end{figure*}

Figure \ref{missing_colors} (a) shows the statistics obtained from
such an experiment. It exhibits the fraction $R_{ge}$ of CA rules
with different alphabet sizes $S$, whose $N$'th supercell version
is missing at least one state. Each data point in this figure was
obtained by testing 10,000 CA rules. The fraction $R_{ge}$
approaches unity at large values of $N$, an expected behavior
since most of the CA are irreversible.

Figure \ref{missing_colors} (b) shows the same data as in (a) when
plotted against the variable $\xi=\tilde{K}\cdot C^S$ where $S$ is
the alphabet size, $\tilde{K}$ is the upper bound for the
Kolmogorov complexity of the supercell CA from Eq.\
(\ref{upper_bound_kc}) and $C$ is a constant. The excellent data
collapse imply a strong correlation between the probability of
finding a missing state and the Kolmogorov complexity of a
supercell CA. This figure also shows that the data points can be
accurately fitted by
\begin{equation}
R_{ge}\left(N,S\right)=\frac{1}{1+\left(\xi/\xi_0\right)^\alpha}\;,
\label{missing_states_scaling_fit}
\end{equation}
with $\xi_0$ a constant and $\alpha\approx 0.7$ (solid line in
Fig.\ \ref{missing_colors} (b)).

Having the scaling form
\begin{eqnarray}
R_{ge}\left(N,S\right)&=&F\left(\xi\right)\;, \\
\xi&=&\tilde{K}\left(N,S\right)\cdot C^S=N^{-1}\cdot
S^{3(1-N)}\cdot C^S\;.\nonumber
\end{eqnarray}
we can now study the behavior of $R_{ge}$ with large alphabet
sizes. Assuming $F$ and $\xi$ to be continuous we define $\xi_h$
as the point where $F\left(\xi_h\right)=1/2$. For a fixed value of
$S$, the slope of $R_{ge}$ at the transition region can be
calculated by
\begin{eqnarray}
\lefteqn{\left.\frac{\partial R_{ge}}{\partial
N}\right|_{N(\xi_h)}=F'\left(\xi_h\right)\cdot
\left.\frac{\partial \xi}{\partial
N}\right|_{N(\xi_h)}=} \nonumber \\
&&-F'\left(\xi_h\right)\cdot\left(N(\xi_h)^{-1}+3\log{S}\right)\cdot\xi_h\;,
\label{Rge_slope}
\end{eqnarray}
where
\begin{eqnarray}
N(\xi_h)&=&\frac{3\log{S}+S\cdot\log{C}-\log{\xi_h}-\log{N(\xi_h)}}{3\cdot\log{S}} \nonumber \\
&\sim&
\frac{S}{\log{S}}\;. \label{Nh}
\end{eqnarray}
Putting together Eqs.\ (\ref{Rge_slope}) and (\ref{Nh}) we find
that the slope of $R_{ge}$ at the transition region grows as
$\log{S}$ for large values of $S$. An indication of this phenomena
can be seen in Fig.\ \ref{missing_colors} (a) which shows sharper
transitions at large values of $S$. In the limit of large $S$,
$R_{ge}$ becomes a step function with respect to $N$. This fact
introduces a critical value $N_c(S)$ such that for $N<N_c(S)$ the
probability of finding a missing state is zero and for $N\geq
N_c(S)$ the probability is one. The value of this critical $N$
grows with the alphabet size as $N_c(S)\sim S/\log{S}$. Note that
$N_c(S)$ is an emergent length scale, as it is not present in any of
the CA rules, but according to the above analysis will emerge (with
probability one) in their dynamics. A direct consequence of the
emergence of $N_c$ is that a measure 1 of all CA can be
coarse-grained to the elementary rule \lq\lq 0" on the
coarse-grained scale $N_c$.

\subsection{Projection probability of CA rules with bounded
Kolmogorov complexity} \label{proj_prob_with_bounded_k}
Generalized
\lq\lq Garden of Eden" states are a specific form of emergent pattern
that can be encountered in the large scale dynamics of CA. Is the
Kolmogorov complexity of CA rules related to other types of
coarse-grained behavior? To explore this question we attempted to project
(solve Eq.\ (\ref{cg_matrix_form})) random CA with bounded Kolmogorov
complexities.

To generate a random CA $A=(a(t),S,f_A)$ with a bounded Kolmogorov
complexity we view the update rule $f_A$ as a string of
$S^3\log_2{S}$ bits, denote the $i$'th bit by $(f_A)_i$ and apply
the following procedure: 1) Randomly pick the first $l$ bits of
$f_A$. 2) Randomly pick a generating function
$G:\{0,1\}^l\rightarrow\{0,1\}$. 3) Set the values of all the
empty bits of $f_A$ by applying $G$:
\begin{equation}
(f_A)_i=G\left[(f_A)_{i-l},(f_A)_{i-l+1},\cdots,(f_A)_{i-1}\right]\;,
\end{equation}
starting at $i=l+1$ and finishing at $i=S^3\log_s{S}$. Up to an
additive constant, the length of such a procedure is equal to
$l+2^l$, the number of random bits chosen. The Kolmogorov
complexity of the resulting rule table can therefore be bounded by
\begin{equation}
K\left(f_A\right)\leq
\bar{K}\left(f_A\right)=\frac{l+2^l}{S^3\log_2{S}}\;.
\label{projection_exp_upper_bound_kc}
\end{equation}
For small values of $l$ this is a reasonable upper bound. However
for large values of $l$ this upper bound is obviously not tight
since the size of $G$ can be much larger than the length of $f_A$.

\begin{figure*}[ht]
\centerline{ \epsfxsize=75mm \epsffile{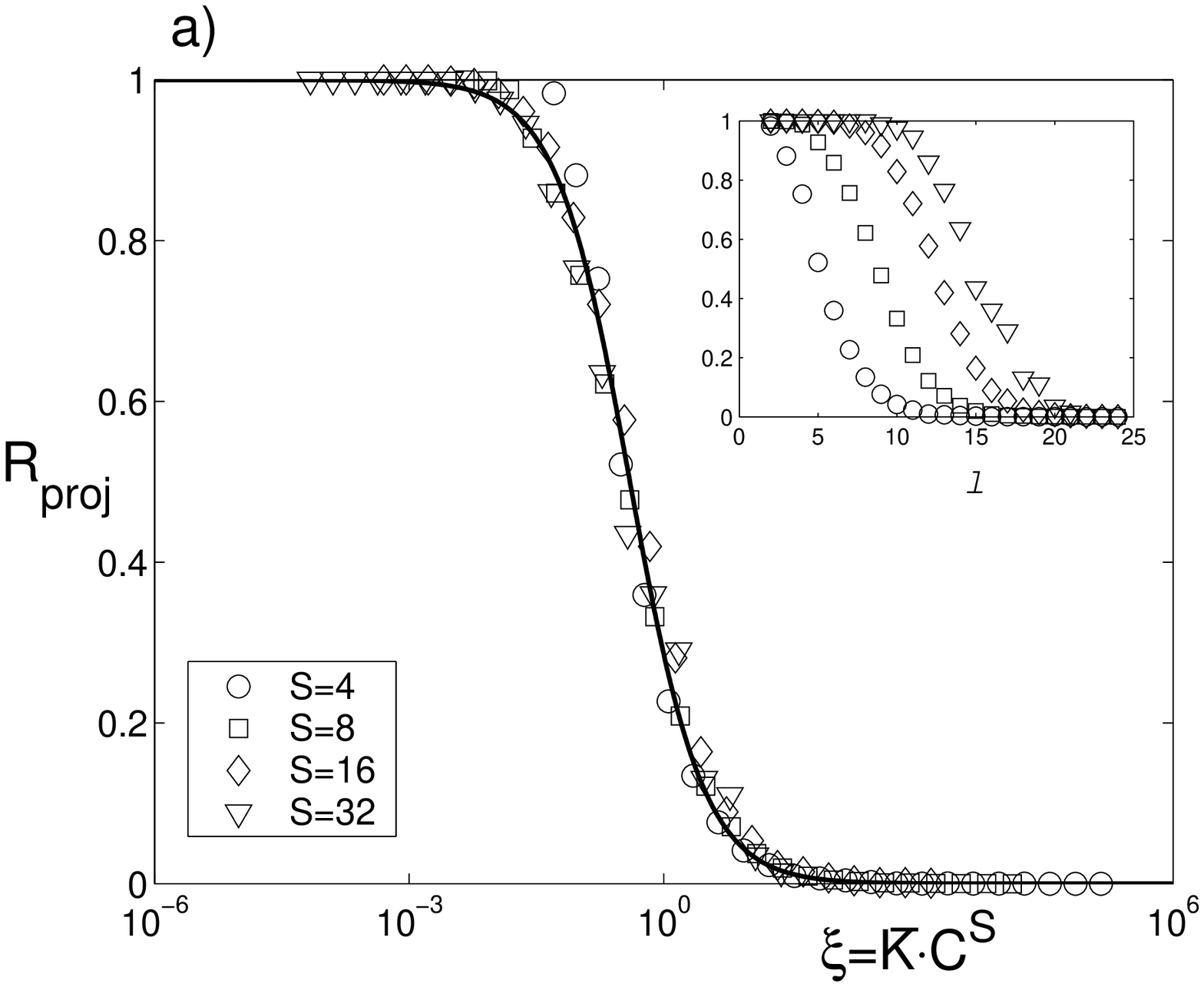}
\epsfxsize=75mm \epsffile{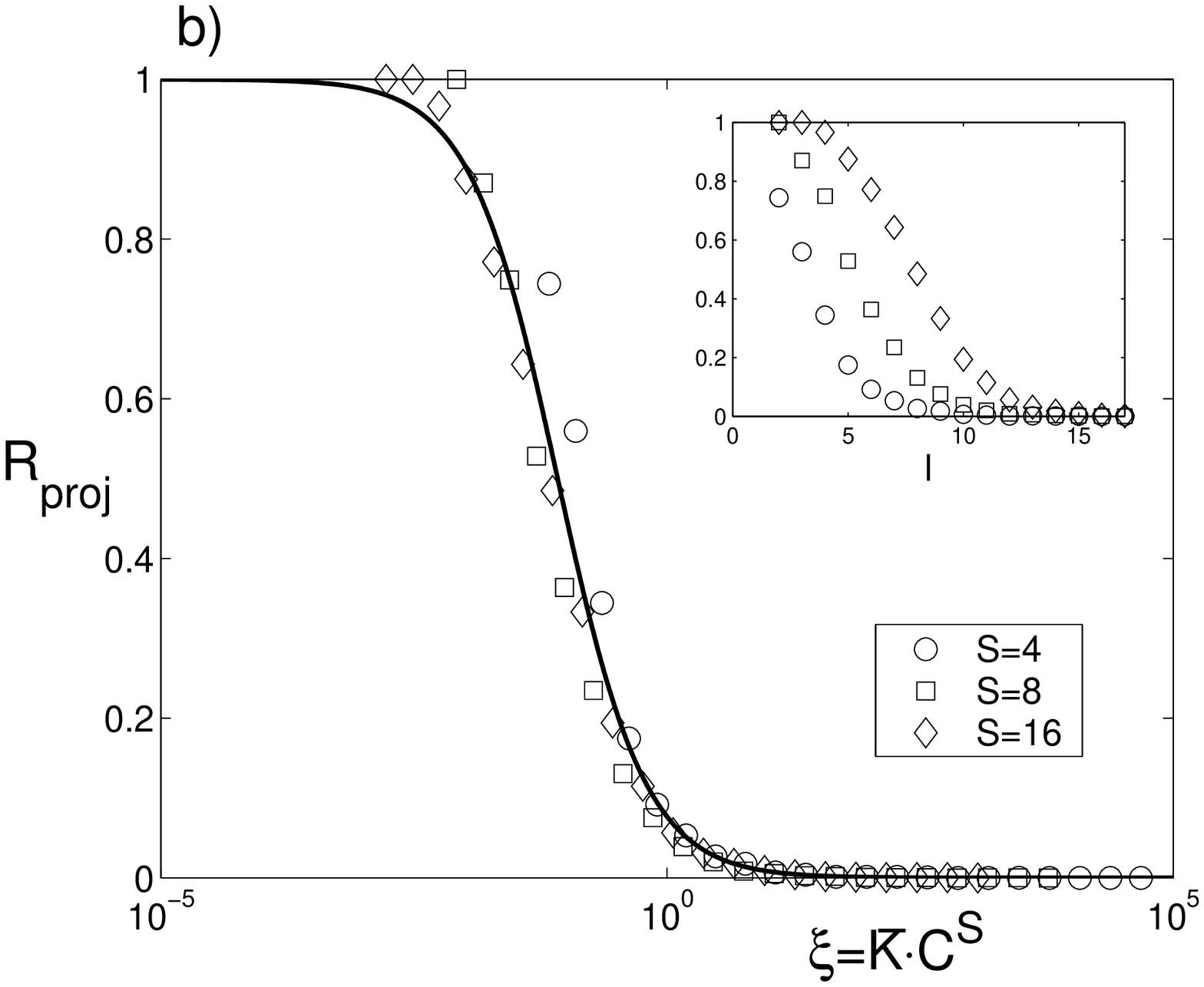}}
\centerline{ \epsfxsize=75mm \epsffile{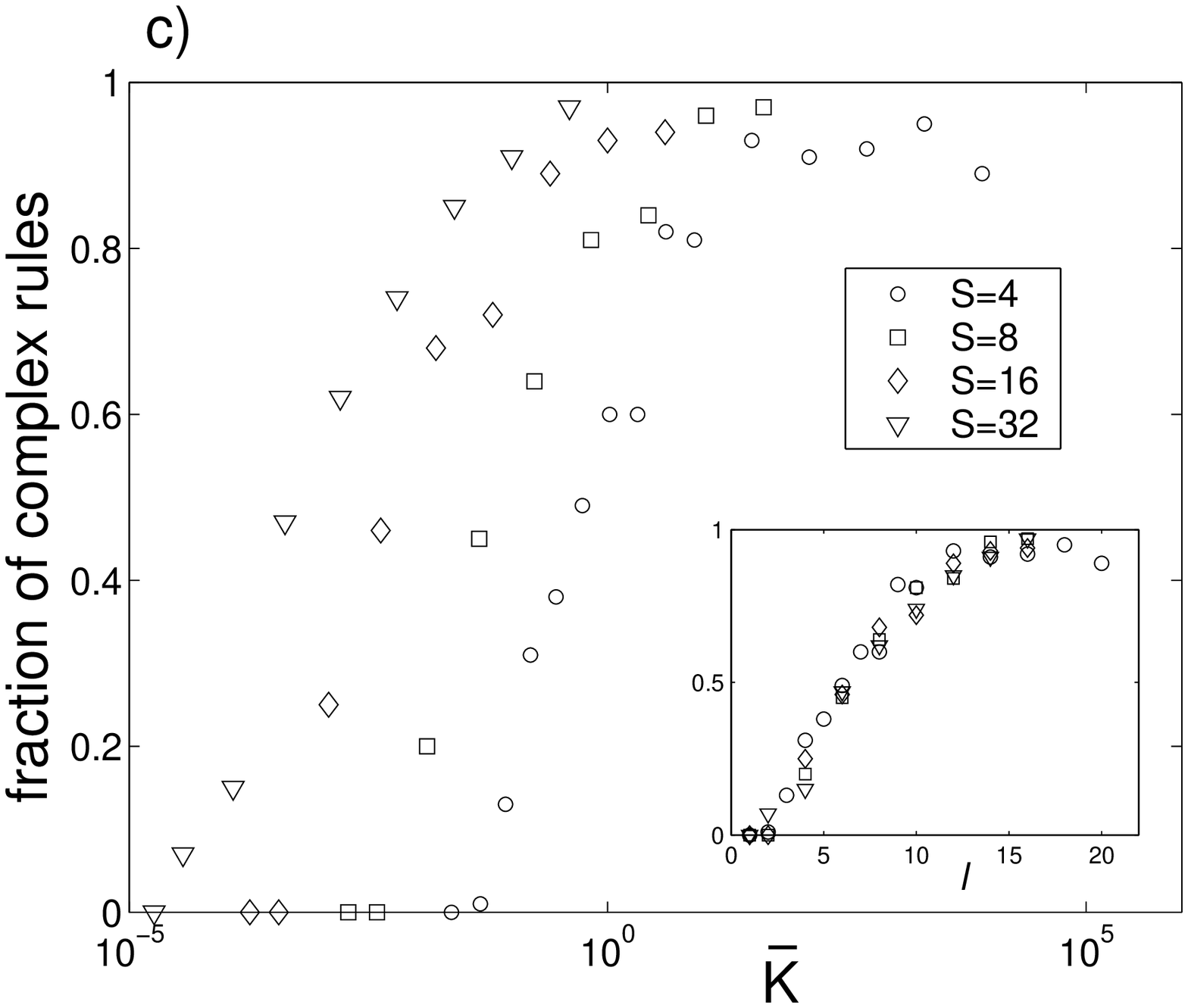}}
\caption{(a) and (b) show the fraction $R_{proj}$ of Kolmogorov
complexity bounded CA that has a valid projection on the binary
alphabet. CA were generated using a random generating function
with $l$ variables according to the procedure described above.
$\bar{K}$ (Eq.\ (\ref{projection_exp_upper_bound_kc})) is the
resulting upper bound Kolmogorov complexity. Different symbols
correspond to different alphabet sizes. Insets show the data as a
function of the parameter $l$. (a) shows results in the case where
all projections $P:S^3\rightarrow\{0,1\}$ are allowed. (b) shows
results in the case where only equal partition projections (Eq.\
(\ref{equal_partition_proj})) are allowed. Solid lines in (a) and
(b) shows a fit by Eq.\ (\ref{Rproj_fit}). (c) shows the fraction
of complex behaving rules which are produced by our procedure as a
function of $l$. } \label{projection_probability}
\end{figure*}

Using the above procedure we studied the probability of projecting
CA with different alphabets and different upper bound Kolmogorov
complexities $\bar{K}$. For given values of $S$ and $l$ we
generated 10,000 (200 for the $S=32$ case) CA and tried to find a
valid projection on the $\{0,1\}$ alphabet. Figure
\ref{projection_probability} (a) shows the fraction $R_{proj}$ of
solvable instances as a function of $\xi=\bar{K}\cdot C^S$. The
constant $C$ used for this data collapse is $1.02$, very close to
$1$. As valid projection solutions we considered all possible
projections $P:S^3\rightarrow\{0,1\}$. In doing so we may be
redoing the missing states experiment because many low Kolmogorov
complexity rules has missing states and can thus be trivially
projected. In order to exclude this option we repeated the same
experiment while restricting the family of allowed projections to
be equal partitions of $\{0\cdots S^3-1\}$, i.e.
\begin{equation}
P:S^3\rightarrow\{0,1\}\;;\;\;|\{x:P(x)=0\}|=|\{x:P(x)=1\}|\;.
\label{equal_partition_proj}
\end{equation}
The results are shown if Fig.\ \ref{projection_probability} (b).

It seems that in both cases there is a good correlation between
the Kolmogorov complexity (or its upper bound) of a CA rule and
the probability of finding a valid projection. In particular, the
fraction of solvable instances goes to one at the low $\bar{K}$
limit. As shown by the solid lines in Fig.\
\ref{projection_probability}, this fraction can again be fitted by
\begin{equation}
R_{proj}=\frac{1}{1+\left(\xi/\xi_0\right)^\alpha}\;,
\label{Rproj_fit}
\end{equation}
where $\xi_0$ is a constant and in this case $\alpha\approx 1$.

How many of the CA rules that we generate and project show a
complex behavior? Does the fraction of projectable rules simply
reflect the fraction of simple behaving rules? To answer this
question we studied the rules generated by our procedure. For each
value of $S$ and $l$ we generated 100 rules and counted the number
of rules exhibiting complex behavior. A rule was labelled \lq\lq
complex" if it showed class 3 or 4 behavior and exhibited a
complex sensitivity to perturbations in the initial conditions.
Fig.\ \ref{projection_probability} (c) shows the statistics we
obtained with different alphabet sizes as a function of $\bar{K}$
while the inset shows it as a function of $l$. We first note that
our statistics support Dubacq et al. \cite{dubacq01}, who
proposed that rule tables with low Kolmogorov complexities lead to
simple behavior and rule tables with large Kolmogorov complexity
lead to complex behavior. Moreover, our results show that the
fraction of complex rules does not depend on the alphabet size and
is only a function of $l$. Rules with larger alphabets show
complex behavior at a lower value of $\bar{K}$. As a consequence,
a large fraction of projectable rules are complex and this
fraction grows with the alphabet size $S$.

As we explained earlier, the Kolmogorov complexity of supercell
versions of CA approaches zero as the supercell size $N$ is
increased. Our experiments therefore indicate that a measure one
of all CA are coarse-grained-able if we use a coarse enough scale.
Moreover, the data collapse that we obtain and the sharp
transition of the scaling function suggest that it may be possible
to know in advance at what length scales to look for valid
projections. This can be very useful when attempting to
coarse-grain CA or other dynamical systems because it can narrow
down the search domain. As in the case of \lq\lq Garden of Eden"
states that we studied earlier, we interpret the transition point
as an emergent scale which above it we are likely to find self
organized patterns. Note however that this scale is a little
shifted in Fig.\ \ref{projection_probability} (b) when compared
with Fig.\ \ref{projection_probability} (a). The emergence scale
is thus sensitive to the types of large scale patterns we are
looking for.

\section{Summary and discussion}
\label{conclusions} In this work we studied emergent phenomena in
complex systems and the associated predictability problems by
attempting to coarse-grain CA. We found that many elementary CA
can be coarse-grained in space and time and that in some cases
complex, undecidable CA can be coarse-grained to decidable and
predictable CA. We conclude from this fact that undecidability and
computational irreducibility are not good measures for physical
complexity. Physical complexity, as opposed to computational
complexity should address the interesting, physically relevant,
coarse-grained degrees of freedom. These coarse-grained degrees of
freedom maybe simple and predictable even when the microscopic
behavior is very complex.

The above definition of physical complexity brings about the
question of the objectivity of macroscopic descriptions
\cite{schulman01,shalizi04}. Is our choice of a coarse-grained
description (and its consequent complexity) subjective or is
it dictated by the system? Our results are in accordance with
Shalizi and Moore \cite{shalizi04}: it is both. In many cases we
discovered that a particular CA can undergo different
coarse-graining transitions using different projection operators.
In these cases the system dictates a set of valid projection
operators and we are restricted to choose our coarse-grained
description from this set. We do however have some freedom to
manifest our subjective interest.

The coarse-graining transitions that we found induce a hierarchy on the
family of elementary CA (see Fig.\ \ref{mapfigure}). Moreover, it seems
that rule complexity never increases with coarse-graining transitions.
The coarse-graining hierarchy therefore provides a partial complexity
order of CA where complex rules are found at the top of the hierarchy
and simple rules are at the bottom. The order is partial because we
cannot relate rules which are not connected by coarse-graining
transitions. This coarse-graining hierarchy can be used as a new
classification scheme of CA. Unlike Wolfram's, classification this
scheme is not a topological one since the basis of our suggested
classification is not the CA trajectories. Nor is this scheme
parametric, such as Langton's $\lambda$ parameter scheme. Our scheme reflects
similarities in the algebraic properties of CA rules. It simply says
that if some coarse-grained aspects of rule $A$ can be captured by the
detailed dynamics of rule $B$ then rule $A$ is at least as complex as
rule $B$. Rule $A$ maybe more complex because in some cases it can do
more than its projection. Note that our hierarchy may subdivide
Wolfram's classes. For example rule 128 is higher on the hierarchy than
rule 0. These two rules belong to class 1 but rule 128 can be
coarse-grained to rule 0 and it is clear that an opposite transition
cannot exist. It will be interesting to find out if class 3 and 4 can
also be subdivided.

In the last part of this work we tried to understand why is it
possible to find so many coarse-graining transitions between CA.
At first blush, it seems that coarse-graining transitions should be
rare because finding valid projection operators is an over
constrained problem. This was our initial intuition when we first
attempted to coarse-grain CA. To our surprise we found that many
CA can undergo coarse-graining transitions.

A more careful investigation of the above question suggests that
finding valid projection operators is possible because of the
structure of the rules which govern the large scale dynamics.
These large scale rules are update functions for supercells, whose
tables can be computed directly from the single cell update
function. They thus contain the same amount of information as the
single cell rule. Their size however grows with the supercell size
and therefore they have vanishing Kolmogorov Complexities.

In other words, the large scale update functions are highly
structured objects. They contain many regularities which can be
used for finding valid projection operators. We did not give a
formal proof for this statement but provided a strong experimental
evidence. In our experiments we discovered that the probability to
find a valid projection is a universal function of the Kolmogorov
Complexity of the supercell update rule. This universal
probability function varies from zero at large Kolmogorov
Complexity (small supercells) to  one at small Kolmogorov
Complexity (large supercells). It is therefore very likely that we
find many coarse-graining transitions when we go to large enough
scales.

Our interpretation of the above results is that of emergence. When we
go to large enough scales we are likely to find dynamically
identifiable large scale patterns. These patterns are emergent (or self
organized) because they do not explicitly exist in the original single
cell rules. The large scale patterns are forced upon the system by the
lack of information. Namely, the system (the update rule, not the cell
lattice) does not contain enough information to be complex at large
scales.

Finding a projection operator is one specific type of an over
constrained problem. Motivated by our results we looked into other
types of over constrained problems. The
satisfyability\cite{kirkpatrick94,monasson99} problem (k-sat) is a
generalized (NP complete) form of constraint satisfaction system. We
generated random 3-sat instances with different number of variables
deep in the un-sat region of parameter space. The generated instances
however were not completely random and were generated by generating
functions. The generating functions controlled the instance's
Kolmogorov complexity, in the same way that we used in section
\ref{proj_prob_with_bounded_k}. We found\cite{unpub_note} that the
probability for these instances to be satisfiable obeys the same
universal probability function of Eq.\ (\ref{Rproj_fit}). It will be
interesting to understand the origin of this universality and its
implications.

In this work, we have restricted ourselves to deal with CA because it
is relatively easy to look for valid projection operators for them. A
greater (and more practical) challenge will now be to try and
coarse-grain more sophisticated dynamical systems such as probabilistic
CA, coupled maps and partial differential equations. These types of
systems are among the main work horses of scientific modelling, and
being able to coarse-grain them will be very useful, and is a topic of
current research, e.g. in material science\cite{GOLD05}. It will be
interesting to see if one can derive an emergence length scale for
those systems like the one we found for \lq\lq Garden of Eden"
sequences in CA (section \ref{GardensOfEden}). Such an emergence length
scale can assist in finding valid projection operators by narrowing the
search to a particular scale.

\begin{acknowledgments}

NG wishes to thank Stephen Wolfram for numerous useful discussions and
his encouragement of this research project. NI wishes to thank David Mukamel for his help and advice. This work was partially
supported by the National Science Foundation through grant
NSF-DMR-99-70690 (NG) and by the National Aeronautics and Space
Administration through grant NAG8-1657.

\end{acknowledgments}

\bibliographystyle{apsrev}
%\bibliography{coarsening}

\begin{thebibliography}{52}
\bibitem[{ent()}]{entropy_comment}
\bibinfo{note}{Another measure for information content is entropy. Entropy
  however is more suitable for ensembles and here we need to quantify the
  information content of individual objects.}
\end{thebibliography}

\end{document}